\documentclass[oneside,english]{amsart}
\usepackage[T1]{fontenc}
\usepackage[latin9]{inputenc}
\usepackage{float}
\usepackage{textcomp}
\usepackage{amsthm}
\usepackage{amstext}
\usepackage{amssymb}
\usepackage{graphicx}
\usepackage{esint}
\usepackage[authoryear]{natbib}

\makeatletter

\providecommand{\tabularnewline}{\\}

\numberwithin{equation}{section}
\numberwithin{figure}{section}

\makeatother

\usepackage{babel}
\begin{document}
\title[Multivariate interactions]{Multivariate interactions modeling through their manifestations: low dimensional model building via the Cumulant Generating Function}
\author{Jhan Rodr\'iguez}
\author{Andr\'as B\'ardossy}

\begin{abstract}
Growing dimensionality of data calls for beyond-pairwise interactions
quantification. Measures of multidimensional interactions quantification
are hindered, among others, by two issues: 1. Interpretation difficulties,
2. the curse of dimensionality. We propose to deal with multidimensional
interactions by identifying subject-matter specific \emph{interaction
manifestations} and then building a low-dimensional model that reproduces
as close as possible such manifestations. We argue that an adequate
model building approach is to build the model in the form of a cumulant
generating function, i.e. to use joint cumulants as building blocks.
The whole approach resembles that of probability inversion in the
area of expert knowledge based risk assessment, where a discrimination
is made between ``elicitation'' variables, familiar to the experts,
and ``target'' (or model) variables, consisting of the more abstract
parameters of a mathematical model. A  synthetic example is provided
to illustrate these ideas.
\end{abstract}
\maketitle

\section*{Introduction}

Technological innovation has lead to a world full of data of an increasingly
growing dimension. These data in turn contain information, the extraction
of which is a basic task of Statistics (c.f. \citet{lindsay2004}).
An important type of information is the kind of interdependence among
variables being represented by data. This calls for statistical means
of extracting, quantifying and, if possible, modeling such interdependence.
At the very least, coefficients that somehow summarize the type and
intensity of multivariate interdependence are very desirable in applied
science.

The introduction of the correlation concept by Francis Galton (1822-1911)
had a tremendous impact on many sciences due to its straightforward
interpretation as a measure of ``partial causation'' or ``average
association'', as summarized in a single parameter. However elementary
this concept now may seem, it was welcomed as an important scientific
contribution at the end of the XIX century. As \citet{pearson_life_2011}
writes, it did ``open to quantitative analysis wide fields of medical,
psychological and social research {[}...{]}, {[}it{]} was to replace
not only in the minds of many of us the old category of causation,
but deeply to influence our outlook on the universe''. 

Karl Pearson would develop the original correlation coefficient of
Galton into the widely used product-moment correlation coefficient.
Given the first paradigmatic step, new implementations of the concept
would appear, in the form of other association coefficients more adequate
for specific applications in psychology and the social sciences: Spearman's
$\rho$, Kendall's $\tau$, Ginni's $\gamma$, Blonqvist's $q$, etc.
(The reader is referred to \citet{joe_relative_1989} for more coefficients).
These are coefficients intended to represent the degree of association
between two random variables.

Later on, \citet{renyi_measures_1959} would attempt to give some
mathematical rigor to the concept of dependence, providing ``seven
rather natural postulates which should be fulfilled by a suitable
measure of dependence''. R\'enyi's work would be revised by \citet{schweizer_nonparametric_1981},
who made some ``reasonable modifications'' to the postulates, since
they were found to be too restrictive. Additionally, \citet{schweizer_nonparametric_1981}
used the concept of copulas to introduce a number of measures of pair-wise
dependence which fulfilled their new postulates. With these conceptual
tools (i.e. a set of reasonable postulates and the unifying concept
of copulas), \citet{wolff1980n} extends the measures of dependence
between two variables given by \citet{schweizer_nonparametric_1981},
and proposes an extension to more than two variables of Spearman's
$\text{\ensuremath{\rho}}$. This course of action has been further
followed and developed by \citet{jaworski_copula-based_2010}. Indeed,
\citet{jaworski_copula-based_2010} introduce a series of measures
that can be considered as extension to more that two variables of
some of the well-established, pair-wise measures of dependence mentioned
above.

Another course of action, traceable back to \citet{linfoot1957informational},
is to use entropy or mutual information as association coefficient.
\citet{joe_relative_1989} proposes a number of measures of this type
that apply to more than two variables; \citet{pena2007dimensionless}
introduce a measure which adjusts itself to dimension, so as to compare
the intensity of association of two vectors of different dimensions.
\citet{Micheas2006765} deal with the general case of $\varphi$-dependence,
of which mutual information is one particular case. The intensity
of association is measured by \citet{Micheas2006765} in terms of
the deviance of the joint distribution from the distribution given
by the product of the marginal distributions (the independence case).
The specific definition of deviance depends on the specific selection
of function $\varphi:\left[0,+\infty\right)\rightarrow\mathbb{R}$,
which is continuous and convex, satisfying some basic conditions. 

Apart from their theoretical interest, measures of dependence for
more than two variables are required and sought in applied research.
In the area of \emph{neuronal science}, an influential theory of behavior
introduced by \citet{hebb_organisation_1949} suggests that ``fundamental
insight into the nature of neuronal computation requires the understanding
of the cooperative dynamics of populations of neurons'' (\citet{grun_analysis_2010},
chapter 12), and further evidence in the course of the years has lead
brain theorists to build models that ``rely on groups of neurons,
rather than single nerve cells, as the functional\emph{ }building
blocks for representation and processing of information'' (\citet{grun_analysis_2010},
preface); this has lead to the development of techniques to quantify
beyond pair-wise association in that research area. Concerning applied
\emph{atmospheric research}, \citet{bardossy_copula_2009} in the
context of daily precipitation modeling, and \citet{bardossy_multiscale_2012}
in the context of downscaling, have found evidence that explicit quantification
of interactions among more than two variables, and their proper incorporation
into modeling and forecasting, may be of an importance hitherto unexplored:
predictions based on statistical models can be otherwise severely
biased, particularly for very high (or extreme) values of the multivariate
process modeled. More recently, \citet{ellipticalSpatialRodriguezBardossy}
investigate the consequences for inference of ignoring multivariate
interdependence in the context of Spatial Statistics, and propose
a model that can deal with this type of interdependence explicitly.
The present paper comprises the theoretical basis for the work of
\citet{ellipticalSpatialRodriguezBardossy}. In the area of \emph{finance},
``herding'' behavior, the degree to which several economic actors
behave as a herd (\citet{Dhaene2012357}), doing basically the same
thing, is important for estimation of loss risks: If a single underlying
factor or small number of factors are inducing a high degree of herd
behavior, financial assets practically independent or very loosely
correlated with each other can interact \emph{en bloc}, rendering
portfolio diversification ineffective. As \citet{Dhaene2012357} indicate,
pair-wise correlations, or a measure based on these, may be misleading
in this case. \citet{RePEc:ner:leuven:urn:hdl:123456789/410070} present
a related interdependence measure for aggregating risks. 

In a recent paper, \citet{reimherr2013} note that most of the theory
on measures of association has left out the important issue of interpretability
of the measures for the research at hand. These authors argue (correctly,
in our opinion) that the lack of interpretability limits the their
use as summary tools. This problem is greatly exacerbated if what
we intend to quantify or represent is the association among several
variables. In a more general manner, the interpretability of parameters
and coefficients of a statistical model has been considered an important
characteristic of the model by \citet{cox1995relation}. 

This paper deals with some of the issues inherent in formulating interaction
coefficients that pertain to more than two variables. We propose an
approach for dealing with these issues. Section \ref{sec:Difficulties-of-defining}
introduces some issues that one encounters when dealing with measures
of interaction for more than two variables. Section \ref{sec:Interaction-parameters-versus-manifestations}
states the approach we suggest for dealing with these issues: to discriminate
between interaction ``parameters'' and interaction ``manifestations''.
We illustrate what we mean by the names \emph{interaction parameter}
and \emph{interaction manifestation}. Section \ref{sec:The-Lancaster-Interaction}
introduces joint cumulants and Lancaster Interactions. The relation
between the two is exhibited, and a justification of joint cumulants
as legitimate extensions to covariance coefficients indicated. Section
\ref{sec:Interaction-manifestations-in-terms} exhibits the relation
between joint cumulants and some illustrative interaction manifestations,
as defined in this paper. Section \ref{sec:Illustration:-Runoff-to}
illustrates the ideas presented, in that a specific model is introduced,
and the ideas of this paper applied to simulated data. In section
\ref{sec:Discussion}, a discussion of the results is provided.

\section{\label{sec:Difficulties-of-defining}Difficulties of defining a measure
of multivariate interaction}

\subsection{\label{sub:Interpretability}Interpretability }

For a two dimensional dataset, interpretability of a dependence coefficient
is aided by the possibility of plotting the data. One looks at several
datasets and computes the respective coefficient of dependence. After
many such data sets, one has an idea of what, say, a correlation coefficient
with a value of $-0.8$ stands for. This visual aid is still possible
for three dimensional datasets, but is not available for higher dimensions.
Assuming we have a coefficient of interdependence, $\lambda$, applicable
to multivariate vectors; how is one supposed to interpret a value
of $\lambda\left(X_{1},X_{2},X_{3},X_{4}\right)=-0.8$? Can one visualize
a dataset producing such a coefficient, so as to relate it to the
phenomenon one is investigating? 

It has been claimed that major advances in the science of statistics
usually occur as a result of the theory-practice interaction (\citet{box1976}),
and that the parameters of a model should have clear subject-matter
interpretations (\citet{cox1995relation}). These statements suggest
that interaction parameters as mere abstract constructions will not
find much application in statistical modeling, unless one can ``paraphrase''
their meaning and relate them to the problem at hand. 

Our approach to interaction quantification and modeling consists in
discriminating between interaction \emph{manifestations} and interaction
\emph{parameters}. So, we can focus on quantification and modeling
of what \emph{really interests us} about dependence in data (i.e.
its subject-matter relevant manifestation), while trying to reproduce
such manifestations with as few parameters as possible.

\subsubsection*{Relation to the ``probability inversion'' technique in Probabilistic
Risk Assessment}

An analogous approach has found successful application in the area
of probabilistic risk assessment (\citet{bedford2001probabilistic}).
With the aid of mathematical models, it is often possible to predict
(approximately) what the consequences of a given event may be. This
mathematical model has parameters that ideally should be calibrated
on the basis of past data. However, absence of data for certain events
(e.g. a nuclear accident in a given region) makes the use of expert
knowledge necessary, whereby the model parameters are to be estimated
on the basis of the experience of a group of experts. Experts usually
cannot give an adequate direct evaluation of the joint probability
distribution of the model parameters, or \emph{target variables}.
Hence each expert is asked to express his uncertainty judgments in
terms of \emph{elicitation variables}, i.e. \emph{observable quantities
}within the area of his/her expertise. A target variable-set for the
model is then recovered, such that the elicitation variables produced
by the mathematical model look as similar as possible like the elicited
variables provided by the experts. This is an inverse problem, labeled
``probabilistic inversion''. The interested reader is referred for
more details to (\citet{bedford2001probabilistic,Du20061164}) and
the references therein.

We suggest in section \ref{sec:Interaction-parameters-versus-manifestations}
a course of action that is analogous to ``probabilistic inversion''
for the problem of interactions quantification and modeling.

\subsection{\label{sub:High-parametric-dimensionality}High parametric dimensionality}

A second issue when defining an interaction coefficient, is the issue
of high dimensionality. As dimension of the random vector under analysis
increases, a naive use of interaction coefficients becomes prohibiting.
For example, the correlation matrix of a 10-dimensional random vector
is an array having $45$ correlation coefficients. Assume symmetry
on the variables with respect to the association coefficient (i.e.
the order of the variables plays no role on the coefficient's value):
If, for the same 10-dimensional vector, one intends to consider 3-wise,
4-wise and 5-wise \textquotedbl{}correlation coefficients\textquotedbl{},
the corresponding arrays would have 450, 4500, and 45000 coefficients,
respectively. 

Hence, it is necessary to be able to select judiciously the interaction
parameters with which to work, and impose reasonable constraints on
them.

Another aspect that can be considered a sort of ``curse'' of dimensionality,
is the coefficient of interdependence to use: there are too many features
that multivariate datasets can exhibit. 

In the one-dimensional case, parameters such as mean, standard deviation,
skewness and kurtosis (basically, the first four cumulants) give a
lot of information about the distribution of data, provided these
data come from an unimodal distribution. Those parameters (mean, skewness
coefficient, etc.) describe data to some extent, since they can be
readily connected to specific questions about data: the location of
data, how informative this location about data is, how symmetric the
distribution is, to what extend can one expect values very far away
from the mean. As a reference one may have in mind these characteristics
for the normal distribution. 

But as dimension grows, one must focus on that feature of data interaction
which is most connected with the research questions at hand, rather
than on an abstract dependence coefficient.

\subsubsection{How high dimensionality is dealt with in the realm of Spatial Statistics}

We present now an example of how the issue of high dimensionality
has been addressed in the context of Spatial Statistics. This will
give us a basis method from which to generalize. 

In the area of spatial statistics (see, for example \citet{cressie_statistics_1991,cressie2011statistics,diggle2007model}),
the studied random vector $\mathbf{X}\in\mathbb{R}^{J}$ spans hundreds
or thousands of components, each of which component represents the
value of an environmental process at a given location $j=1,\ldots,J$.
The way high dimensionality is addressed in spatial statistics is
an apt introduction for the method we advocate in this paper. We give
here a very basic form of a spatial statistical model, since it suffices
for our introducing purposes.

One focuses on the correlation between every two components of $\mathbf{X}$.
The covariance among every two components, $\left(X_{i},X_{j}\right)$,
of $\mathbf{X}$, is expressed as a function $Cov\left(d\right)$
of the distance between the locations represented by these two components,
$d\geq0$. The covariance function $Cov\left(d\right)$ must be such
that the resulting covariance matrix is positive definite. To this
end there are a number of covariance functions often used in practice,
for example, one popular covariance function is the powered exponential
one, 
\begin{equation}
Cov\left(d\right)=\sigma_{0}^{2}.I\left(d=0\right)+\sigma_{1}^{2}\exp\left(-\left(d/\theta_{1}\right)^{\theta_{2}}\right)\label{eq:Power-exp_cov}
\end{equation}
where $\theta_{1}>0$, $0<\theta_{2}\leq2$, $\sigma_{0}^{2}\geq0$,
$\sigma_{1}^{2}\geq0$ are the covariance function parameters.

Note that:
\begin{enumerate}
\item Function (\ref{eq:Power-exp_cov}) allows to have the covariance between
every two components of $\mathbf{X}$ as a function of the distance
between the locations these components represent, and only 4 additional
parameters. In this way, the whole dependence structure of $\mathbf{X}\in\mathbb{R}^{J}$
(with $J>>2$) is \emph{low dimensionally} obtained, built on the
basis of 2-dimensional dependence coefficients. 
\item The interesting \emph{dependence manifestation} to recover is covariance
between every two components of $\mathbf{X}$, whereas the (interaction)
parameters to estimate are the function parameters, $\theta_{1},\theta_{2},\sigma_{0}^{2}$
and $\sigma_{1}^{2}$. This is entirely analogous to the probability
inversion technique mentioned in section \ref{sub:Interpretability}:
covariance takes the place of the elicitation variables, whereas the
covariance function parameters are the target variables.
\item There is a functional relation between $\theta_{1},\theta_{2},\sigma_{0}^{2},\sigma_{1}^{2}$
and the dependence manifestation. Covariance can be written in terms
of the (interaction) parameters, $\theta_{1},\theta_{2},\sigma_{0}^{2}$
and $\sigma_{1}^{2}$.
\end{enumerate}
Items 1 through 3 summarize a technique to tackle the problem of high
dimensionality in an ingenious low-dimensional way. The issue of interpretability
goes relatively unnoticed, since in this case parameters have a relatively
straightforward interpretation: $\sigma_{0}^{2}$ represents a micro
scale variability of the environmental process; $\sigma_{0}^{2}+\sigma_{1}^{2}$
represents the variance of the marginal distribution of each component
$X_{j}$ of $\mathbf{X}$; $\theta_{1}$ (often called ``range'')
represent the distance at which correlation between data from two
locations is relatively insignificant. The parameter $\theta_{2}$
might even receive a suitable interpretation, depending on the context.

In the next section, this approach is extended to deal with the interdependence
among more than two variables at a time, keeping basically the same
ideas.

\section{\label{sec:Interaction-parameters-versus-manifestations}Interaction
parameters versus interaction manifestations}

The approach we advocate in this paper can be summarized as follows:
first select an interaction ``manifestation'' relevant for the research
in question. Then fit (low-dimensional) interactions ``parameters''
that make the fitted distribution reproduce, as close as possible,
the observed interaction manifestation. In this way, we circumvent
the issues of interpretability and high dimensionality mentioned above. 

By interaction manifestation, we mean any function of more than one
component of the random vector analyzed, $\mathbf{X}\in\mathbb{R}^{J}$,
which can be interpreted as relevant for the research objectives at
hand. For the sake of illustration:
\begin{enumerate}
\item The distribution of the sum of subsets of components of a random vector.
In the context of financial analysis, this sum is readily interpreted
as ``risk'' (see also section \ref{sec:Illustration:-Runoff-to}
below). 
\item The joint distribution of subsets of components, or the probability
of trespassing simultaneously a threshold defined for each component.
This is useful in many applications. For example, in the context of
series systems reliability, such trespassing probability is the probability
of ``failure''. 
\item Differential entropy, any information-based dependence measure, or
any of the copula-based generalizations to correlation measures studied
by \citet{jaworski_copula-based_2010}, of subsets of components.
Depending on the specific research carried out, these may have subject-matter
interpretations, or can readily provide the versed researcher of a
specific area with a summary picture of the dependence in the data. 
\end{enumerate}
Interaction manifestations are interesting for the problem at hand,
we would like our model to reproduce them properly. But they are not
very helpful for building a model that integrates them, let alone
a low-dimensional model. \emph{If we had }interaction parameters or
coefficients which:
\begin{enumerate}
\item Provide us with an idea of the number of variables interacting within
the random vector analyzed, $\mathbf{X}\in\mathbb{R}^{J}$.
\item Can be somehow (functionally) connected with the interaction manifestations
that are interesting for the research carried out. 
\item Can be built into a parametric or semi-parametric model. This would
immediately open up the possibility of a low-dimensional model, via
a judicious selection of assumptions and/or constraints on the interaction
parameters.
\end{enumerate}
\emph{Then we could} proceed, in the manner of an inverse problem,
as follows:
\begin{enumerate}
\item We find data-based estimates or approximations to the interesting
interaction manifestations
\item We fit the interactions parameters so as to match best the observed
interaction manifestations
\end{enumerate}
In the next section, we introduce a reasonable interaction measure,
and through it, a reasonable type of interaction parameter with which
one can work along the lines above; namely the joint cumulant. We
claim that using joint cumulants as building blocks of a multivariate
statistical model allows for an adequate consideration of dependence,
both of pairs of variables, and of groups of more variables. 

It might be argued that moments (and hence cumulants) of sufficiently
high orders might not exist for the ``true'' probability distribution
of the process under analysis. We would answer that such distributions
can always be sufficiently (i.e. for practical purposes) approximated
by a distribution with existing moments of all orders. See, for example
\citet{1987}, where the authors introduce a semi-parametric model,
similar to an Edgeworth expansion. This model possesses moments of
all orders. Yet, under minimal conditions it can approximate \emph{any}
continuous distribution on $\mathbb{R}^{J}$, provided sufficiently
many factors are added to the sum defining the model. Additionally,
\citet{del_brio_gramcharlier_2009,mauleon2000testing,perote_multivariate_2004}
present variants of the model of \citet{1987}, and show how they
can be effectively applied to modeling heavy tailed data, both univariate
and multivariate.

\section{\label{sec:The-Lancaster-Interaction}The Lancaster Interaction Measure
and Joint Cumulants}

In this section, the connection between the Lancaster Interaction
measure of a random variable and its joint cumulants is established.
To our knowledge, this connection has not been pointed out before
as a justification of joint cumulants as reasonable interdependence
parameters.

\subsection{A review of Lancaster Interactions}

We review now the function called ``additive interaction measure''
or ``Lancaster interaction measure'', introduced by \citet{lancaster_chi-squared_1969}
and later modified by \citet{streitberg_lancaster_1990}. This function
can be built for every random vector $\mathbf{X}\in\mathbb{R}^{J}$,
and has the property of being identically zero if any sub-vector of
$\mathbf{X}$ is independent of the other. 

An additive interaction measure $\Delta F$$\left(\mathbf{X}\right)$
is a signed measure determined by a given distribution $F\left(\mathbf{X}\right)$
on $\mathbb{R}^{J}$. Its defining characteristic is that it is equal
to zero for all $\mathbf{X}\in\mathbb{R}^{J}$, if $F\left(\mathbf{X}\right)$
can be written as the non-trivial product of two or more of its (multivariate)
marginal distributions (\citet{streitberg_lancaster_1990}). For example,
if $J=4$ and $F$ can be written as $F_{124}F_{3}$, being $F_{124}$
and $F_{3}$ the marginal distributions of $\left(X_{1},X_{2},X_{4}\right)$
and $X_{3}$, respectively, then $\Delta F\left(\mathbf{X}\right)\equiv0$,
for all $\mathbf{X}\in\mathbb{R}^{J}$. 

An alternative explanation is that $\Delta F\equiv0$, if one subset
of $\mathbf{X}$'s components is independent of another subset of
components. If $\Delta F\equiv0$, then $F$ is said to be \textquotedbl{}decomposable\textquotedbl{}. 

Lancaster Interaction measure is defined by 
\begin{equation}
\Delta F\left(\mathbf{X}\right)=\sum_{\pi}\left\{ \left(\left(-1\right)^{\left|\pi\right|-1}\left(|\pi|-1\right)!\right)F_{\pi}\left(\mathbf{X}\right)\right\} \label{eq:Lancaster_measure_definition}
\end{equation}
where the sum is over all partitions, $\pi$, of index set $C=\left\{ 1,\ldots,J\right\} $. 

An example will help clarify the notation: for index set $C=\left\{ 1,2,3,4\right\} $
there are 15 partitions, three of which are: $\pi_{1}=\left\{ \left\{ 1\right\} ,\left\{ 2\right\} ,\left\{ 3,4\right\} \right\} $,
$\pi_{2}=\left\{ \left\{ 1,4\right\} ,\left\{ 2,3\right\} \right\} $,
$\pi_{3}=\left\{ \left\{ 1,2,3,4\right\} \right\} $. Their cardinalities
are $\left|\pi_{1}\right|=3$ , $\left|\pi_{2}\right|=2$ and $\left|\pi_{3}\right|=1$,
respectively. In general, a set of $J$ elements has a total of $B_{J}$
possible partitions%
\footnote{The number $B_{J}$ is often called Bell's number.%
}, where $B_{0}=B_{1}=1$ and any subsequent $B_{k>1}$ can be found
(see e.g. \citet{rota_number_1964}) by the recurrence relation $B_{k+1}=\sum_{r=0}^{k}{k \choose r}B_{r}$.
The reader is referred to the textbook of \citet{AignerDiskrete}
for more on partitions and their enumeration.

The symbol $F_{\pi_{1}}$ is further to be interpreted as 
\begin{equation}
F_{\pi_{1}}\left(\mathbf{X}\right)=F_{1}\left(X_{1}\right)F_{2}\left(X_{2}\right)F_{34}\left(X_{3},X_{4}\right)\label{eq:FActorized_distribution_example}
\end{equation}
that is, the product of the (multivariate) marginal distributions
defined by partition $\pi_{1}$. The same explanation holds at (\ref{eq:Lancaster_measure_definition})
for any of the $B_{J}$ partitions, $\pi$, of index set $C=\left\{ 1,\ldots,J\right\} $. 

It will be convenient to define partition operator $J_{\pi}$, to
be applied to $F$ for a given partition $\pi$, by 
\begin{equation}
J_{\pi}F\rightarrow F_{\pi}\label{eq:partition_operator}
\end{equation}
where $F_{\pi}$ is as in the example at equation (\ref{eq:FActorized_distribution_example}). 

\citet{streitberg_lancaster_1990,streitberg_alternative_1999} shows
an important result concerning $\Delta F$: given a probability distribution
function $F$, function $\Delta F$ as in (\ref{eq:Lancaster_measure_definition})
is the \emph{only} function built as a linear combination of products
of (multivariate) marginal distributions of $F$, such that $\Delta F\left(\mathbf{X}\right):=0$,
whenever one subset of $\mathbf{X}$'s components is independent of
another components subset. 

Since the interaction measure is defined in terms of a given distribution
$F$, we can define the interaction operator:
\begin{equation}
\Delta=\sum_{\pi}\left\{ \left(\left(-1\right)^{\left|\pi\right|-1}\left(\left|\pi\right|-1\right)!\right)J_{\pi}\right\} \label{eq:interactions_operator}
\end{equation}
which, upon application to the distribution in question, returns the
additive interaction measure.

\subsection{A review of Joint Cumulants}

Moments and cumulants can be defined as constants summarizing important
information about a probability distribution and sometimes, even determining
it completely (cf. \citet{kendall_advanced_1969}). In this section
we deal with random variables having a probability density function.
The development is also valid for discreet distributions, under simple
modifications. The reader is referred to \citet{kendall_advanced_1969,muirhead_aspects_1982,billingsley_probability_1986,mccullagh_tensor_1987}
for more details on moments and cumulants. 

The Cumulant Generating Function (c.g.f.), $K_{\mathbf{X}}\left(\mathbf{t}\right)$,
of a random vector, $\mathbf{X}\in\mathbb{R}^{J}$, is defined as
the logarithm of the moment generating function (m.g.f.),
\begin{equation}
K_{\mathbf{X}}\left(\mathbf{t}\right)=\log\left(M_{\mathbf{X}}\left(\mathbf{t}\right)\right)=E\left(\exp\left(\sum_{j=1}^{J}t_{j}X_{j}\right)\right)
\end{equation}
where $\mathbf{t}\in\mathbb{R}^{J}$, assuming these functions exist.

Joint cumulants are then defined to be the coefficients of the Taylor
expansion for $K_{\mathbf{X}}\left(\mathbf{t}\right)$, 
\begin{equation}
K_{\mathbf{X}}\left(\mathbf{t}\right)\sim\sum_{r_{1=0}}^{\infty}\ldots\sum_{r_{J}=0}^{\infty}\frac{\kappa_{r_{1},\ldots,r_{J}}.t_{1}^{r_{1}}\ldots t_{J}^{r_{J}}}{r_{1}!\ldots r_{J}!}
\end{equation}
and hence can be found by differentiating $K_{\mathbf{X}}\left(\mathbf{t}\right)$
and evaluating at $\mathbf{t}=\mathbf{0}$,
\begin{equation}
\kappa_{r_{1},\ldots,r_{J}}=\frac{\partial^{r_{1}+\ldots+r_{J}}}{\partial^{r_{J}}t_{J}\ldots\partial^{r_{1}}t_{1}}K_{\mathbf{X}}\left(\mathbf{t}\right)\mid_{\mathbf{t}=\mathbf{0}}
\end{equation}
where $r_{j}\geq0$ is a non-negative integer. An important particular
case is the covariance coefficient, or second order joint cumulant,
\[
\frac{\partial^{2}}{\partial t_{i}\partial t_{j}}K_{\mathbf{X}}\left(t_{i},t_{j}\right)\mid_{\left(t_{i},t_{j}\right)=\left(0,0\right)}=cov\left(X_{i},X_{j}\right)
\]

The c.g.f. of a sub-vector $\mathbf{Y}=\left(X_{j_{1}},\ldots,X_{j_{k}}\right)$,
with indexes in an index set, $j_{i}\in I$, can be readily found
in terms of that of $\mathbf{X}$, by setting the indexes not corresponding
to $\mathbf{Y}$ to zero: 
\begin{multline*}
K_{\mathbf{Y}}\left(\mathbf{s}\right)=\left(E\left(\exp\left(\sum_{i=1}^{k}s_{i}X_{j_{i}}\right)\right)\right)=\log\left(E\left(\exp\left(\sum_{j=1}^{J}g_{j}\left(\mathbf{s}\right)X_{j}\right)\right)\right)=K_{\mathbf{X}}\left(g\left(\mathbf{s}\right)\right)
\end{multline*}
where $g:\mathbb{R}^{k}\rightarrow\mathbb{R}^{J}$, and 
\[
g_{j}\left(\mathbf{s}\right)=\begin{cases}
1, & j\in I\\
0, & j\notin I
\end{cases}
\]

An alternative definition for joint cumulants uses product moments
as departing point (see, for example, \citet{brillinger_time_1974}).
Let $\mathbf{X}\in\mathbb{R}^{J}$ be a random vector. For a set $\left(X_{j_{1}},\ldots,X_{j_{d}}\right)$
of $\mathbf{X}$\textasciiacute{}s components, where some sub-indexes
$j_{r}$ may be repeated, consider joint moments 
\[
E\left(X_{j_{1}}\ldots X_{j_{d}}\right)
\]

Consider partition operator $J_{\pi}^{*}$, analogous to (\ref{eq:partition_operator}),
related to each partition $\pi$ of $\left(j_{1},\ldots,j_{d}\right)$.
This operator converts $E\left(X_{j_{1}}\ldots X_{j_{d}}\right)$
into the product of the factors determined by partition $\pi$. 

For example, for $d=4$ , $\left(j_{1},j_{2},j_{3},j_{3}\right)$
and $\pi=\left\{ \left\{ 1\right\} ,\left\{ 2,3\right\} ,\left\{ 4\right\} \right\} $,
one has partition components $v_{1}=\left\{ 1\right\} $, $v_{2}=\left\{ 2,3\right\} $
and $v_{3}=\left\{ 4\right\} $. Upon application of $J_{\pi}^{*}$,
we have, 
\[
J_{\pi}^{*}E\left(X_{j_{1}}\ldots X_{j_{4}}\right)=E\left(X_{j_{1}}\right)E\left(X_{j_{2}}X_{j_{3}}\right)E\left(X_{j_{3}}\right)
\]

In the general case

\[
J_{\pi}^{*}E\left(X_{j_{1}}\ldots X_{j_{d}}\right)=\prod_{v\in\pi}E\left(\prod_{j_{r}\in v}X_{j_{r}}\right)
\]

The alternative definition of joint cumulants can now be given.

For random variables $\left(X_{j_{1}},\ldots,X_{j_{d}}\right)$, their
joint cumulant of order \emph{d} is given by, 
\begin{multline}
cum\left(X_{j_{1}},\ldots,X_{j_{d}}\right):=\sum_{\pi}\left\{ \left(\left(-1\right)^{\left|\pi\right|-1}\left(\left|\pi\right|-1\right)!\right)J_{\pi}^{*}\right\} E\left(X_{j_{1}}\ldots X_{j_{d}}\right)\label{eq:Joint_cumulants_Brillinger}
\end{multline}

Two examples are:

\begin{eqnarray*}
cum\left(X_{1},X_{2}\right) & = & E\left(X_{1}X_{2}\right)-E\left(X_{1}\right)E\left(X_{2}\right)
\end{eqnarray*}
and
\begin{multline*}
cum\left(X_{1},X_{2},X_{3}\right)=E\left(X_{1}X_{2}X_{3}\right)-E\left(X_{1}X_{2}\right)E\left(X_{3}\right)-E\left(X_{1}X_{3}\right)E\left(X_{2}\right)\\
-E\left(X_{2}X_{3}\right)E\left(X_{1}\right)+2E\left(X_{1}\right)E\left(X_{2}\right)E\left(X_{3}\right)
\end{multline*}

Hence joint cumulants can be seen, from a merely formalistic point
of view, to form a kind of higher order covariance coefficient. The
second order joint cumulant is just the typical covariance coefficient.

\subsection{Relationship between Lancaster Interactions and Joint Cumulants}

The similarity between (\ref{eq:Lancaster_measure_definition}) and
(\ref{eq:Joint_cumulants_Brillinger}) is evident. Indeed, if we concentrate
for now on the case $\mathbf{X}\in\mathbb{R}^{2}$, then \citet{lehmann_concepts_1966}
reports that: 
\begin{multline}
Cov\left(X_{1},X_{2}\right)=cum\left(X_{1},X_{2}\right)=\\
\intop_{-\infty}^{+\infty}\intop_{-\infty}^{+\infty}\left[F_{12}\left(x_{1},x_{2}\right)-F_{1}\left(x_{1}\right)F_{2}\left(x_{2}\right)\right]dx_{1}dx_{2}\label{eq:Hoeffding_Formula}
\end{multline}
under the condition that $E\left(\left|X_{1}^{k_{1}}X_{2}^{k_{2}}\right|\right)<+\infty$,
for $k_{j}=0,1$.

This equation is often called \textquotedbl{}Hoeffding's formula\textquotedbl{}
since it was first discovered by \citet{hoeffding_masstabinvariante_1940}.
Of course, the above equation can be written in terms of the Lancaster
interaction measure (\ref{eq:Lancaster_measure_definition}), as
\begin{equation}
cum\left(X_{1},X_{2}\right)=\intop_{-\infty}^{+\infty}\intop_{-\infty}^{+\infty}\Delta F\left(x_{1},x_{2}\right)dx_{1}dx_{2}\label{eq:Hoeffding_Formula2}
\end{equation}

It turns out that this equation can be extended to higher dimensions.
Let $\mathbf{X}\in\mathbb{R}^{J}$ be a random vector. As shown by
\citet{block_multivariate_1988}, we have that (page 1808):
\begin{equation}
cum\left(\mathbf{X}\right)=\left(-1\right)^{J}\intop_{-\infty}^{+\infty}\ldots\intop_{-\infty}^{+\infty}\sum_{\pi}\left\{ \left(\left(-1\right)^{\left|\pi\right|-1}\left(\left|\pi\right|-1\right)!\right)F_{\pi}\right\} d\mathbf{X}\label{eq:cumul_lancaster}
\end{equation}
under the condition that $E\left(\left|X_{j}^{J}\right|\right)<+\infty$,
for $j=1,\ldots,J$. Again, this is the same as saying that

\begin{equation}
cum\left(\mathbf{X}\right)=\left(-1\right)^{J}\intop_{-\infty}^{+\infty}\ldots\intop_{-\infty}^{+\infty}\Delta F\left(\mathbf{X}\right)d\mathbf{X}\label{eq:cumul_lancaster-1}
\end{equation}

Thus, joint cumulants are equal (up to a known constant) to the integral
of Lancaster Interaction measure; they are ``summary'' or ``integral''
measures of additive interaction. To our knowledge, this connection
had not been pointed out elsewhere.

It goes without much explanation that the joint cumulants of a random
vector $\mathbf{X}$ vanish whenever a subset of the vector is independent
of another, since then the integrating function is identically zero.
This property is well-known and oftentimes the reason why joint cumulants
are used in practice (e.g. in \citet{brillinger_time_1974,mendel_tutorial_1991}).
The inverse is true only if the distribution of $\mathbf{X}$ is determined
by its moments, which may or may not be a reasonable assumption, depending
on the application. Again, based on the work of \citet{1987,perote_multivariate_2004,mauleon2000testing,del_brio_gramcharlier_2009},
we argue that this is not an extreme limitation to our approach, since
all we are seeking is a good approximation to the distribution under
analysis. 

In particular, whenever we have $cum\left(X_{j_{1}},\ldots,X_{j_{d}}\right)\neq0$,
where no index $j_{k}$ is repeated, this means that one cannot decompose
the distribution of $\left(X_{j_{1}},\ldots,X_{j_{d}}\right)$: At
least $d$ variables within $\mathbf{X}$ are interacting simultaneously
with each other. 

Our theoretical contribution here is that joint cumulants are seen
as the integral of the Lancaster interaction measure. As shown by
\citet{streitberg_lancaster_1990}, $\Delta F$ is the only additive
measure, built very elementarily with the marginal distributions of
the random vector, which vanishes whenever one subset of $\mathbf{X}$'s
components is independent of another subset of components.

We have provided a theoretical basis for declaring joint cumulants
``interaction parameters'', and the cumulant generating function
a ``dependence structure''. The functional character of the c.g.f.
opens up the possibility of parametric modeling, with its respective
low-dimensionality advantage. It is just another way of defining a
model, alternative to the density specification. 

We shall see below, how the parameters of a model expressed as a c.g.f.
can be connected with some interesting interaction manifestations.

\section{\label{sec:Interaction-manifestations-in-terms}Interaction manifestations
in terms of interaction parameters}

The connection between interaction parameters (i.e. joint cumulants)
and interaction manifestations relies on the concepts of the Edgeworth
expansion and the saddlepoint approximation to the density of a random
vector. A brief review of these topics is provided at the appendix.

\subsection{Connection of dependence structure with interaction manifestations}

We shall show explicitly the connection of joint cumulants and the
c.g.f. with three of the interaction manifestations listed at section
\ref{sec:Interaction-parameters-versus-manifestations}, which manifestations
refer to subsets of components, $\left(X_{j_{1}},\ldots,X_{j_{k}}\right)$,
$1\leq k\leq J$, of the random vector $\mathbf{X}\in\mathbb{R}^{J}$.
Namely: the distribution of the sum of components; parameters related
to the joint probability of the components; and the differential entropy
of the components.

A relevant point here is that, except for the distribution of the
sum of components, even with a lot of data at hand, estimation of
the interaction manifestations mentioned can be done only for (multivariate)
marginals of relatively low dimension, such as $k$ equal to 3, 4
or 5. But armed with a sensible c.g.f., we can consistently integrate
these manifestations into the whole distribution (in much the same
way as thousand of covariance coefficients are integrated into a Spatial
Statistics model that spans thousands of variables). This we can attain
with the aid of the overarching dependence structure, that is, the
c.g.f. 

Assume for the moment you have a reasonable type of c.g.f., that is,
one that seems reasonable for the problem at hand (for an illustration
see section \ref{sec:Illustration:-Runoff-to}).

\subsubsection{\label{sub:Connection-of-dependence-sumas}Connection of dependence
structure with Sums of components}

Given a random vector $\mathbf{X}\in\mathbb{R}^{J}$ representing
the variables under analysis, we are interested in the distribution
of variable $S_{\mathbf{X}}=\sum_{i=1}^{k}X_{j_{i}}$, where $\left(X_{j_{1}},\ldots,X_{j_{k}}\right)$,
$1\leq k\leq J$, is a sub-vector of the random vector $\mathbf{X}\in\mathbb{R}^{J}$.
The distribution of $S_{\mathbf{X}}$ is the interaction manifestation
we in which we are interested. We want to fit the distribution of
the whole vector, $\mathbf{X}\in\mathbb{R}^{J}$, in such a way the
we fit this interaction manifestation properly.

\emph{One course of action} is to find the cumulants of $S_{\mathbf{X}}$
in terms of the joint cumulants of $\mathbf{X}$, and then approximate
the density of $S_{\mathbf{X}}$, by using the Edgeworth Expansion.
Since $S_{\mathbf{X}}$ is a one-dimensional random variable, one
can alternatively find research-relevant quantiles of its distribution
by inverting the Edgeworth Expansion, i.e. by using the Cornish-Fisher
inversion. 

To find the cumulants of $S_{\mathbf{X}}$, note that two of the properties
of joint cumulants are \citet{brillinger_time_1974}: symmetry and
multi-linearity. Symmetry means that $cum\left(X_{j_{1}},\ldots,X_{j_{k}}\right)=cum\left(P\left(X_{j_{1}},\ldots,X_{j_{k}}\right)\right)$
for any permutation $P\left(j_{1},\ldots,j_{k}\right)$ of the indexes
$\left(j_{1},\ldots,j_{k}\right)$. Concerning multi-linearity, for
any random variable $Z\in\mathbb{R}$, one has
\[
cum\left(Z+X_{j_{1}},\ldots,X_{j_{k}}\right)=cum\left(Z,\ldots,X_{j_{k}}\right)+cum\left(X_{j_{1}},\ldots,X_{j_{k}}\right)
\]
 Combining these two properties, it can be shown that
\begin{equation}
\kappa_{r}\left(S_{\mathbf{X}}\right)=cum\left(\underbrace{S_{\mathbf{X}},\ldots,S_{\mathbf{X}}}_{r}\right)=\sum_{i_{1}=1}^{k}\left[\sum_{i_{2}=1}^{k}\ldots\left[\sum_{i_{r}=1}^{k}cum\left(X_{j_{i_{1}}},\ldots,X_{j_{i_{r}}}\right)\right]\right]\label{eq:cumulants-of-sums}
\end{equation}
where $\kappa_{r}\left(S_{\mathbf{X}}\right)$ denotes the \emph{r}-th
cumulant of random variable $S_{\mathbf{X}}=\sum_{i=1}^{k}X_{j_{i}}$.
Then the interesting quantiles of $S_{\mathbf{X}}$ can be (approximately)
written in terms of the $\kappa_{r}$ via the Cornish-Fisher inversion. 

As the dimension $k$ of the sub-vector increases, this approach becomes
impractical, since the sum at (\ref{eq:cumulants-of-sums}) comprises
too many elements. Fortunately, knowing the c.g.f. of $\mathbf{X}$
tells much about the c.g.f. of sums of its components.

\emph{A second course of action} uses all the information provided
by the c.g.f. and is now given.

In a somewhat more general context as before, consider a random vector
$\mathbf{X}=\left(X_{1},\ldots,X_{J}\right)$. One wishes to study
the joint distribution of aggregated variables of the form:
\begin{eqnarray}
\xi_{1} & = & \sum_{j_{1}\in I_{1}}X_{j_{1}}\nonumber \\
\xi_{2} & = & \sum_{j_{2}\in I_{2}}X_{j_{2}}\\
\vdots & \vdots & \vdots\nonumber \\
\xi_{l} & = & \sum_{j_{l}\in I_{l}}X_{j_{l}}
\end{eqnarray}

where $I_{k}$, for $k=1,\ldots,l$ represent non-overlapping index
sets such that

\[
I_{1}\cup\ldots\cup I_{l}=\left\{ 1,\ldots,J\right\} 
\]
(Note that $S_{\mathbf{X}}$ above is the specific case in which $I_{1}=\left\{ 1,\ldots,J\right\} $).

The cumulant generating function of the $l$-dimensional vector so
obtained is given by
\begin{multline}
K_{\mathbf{\xi}}\left(\mathbf{t}\right)=\log\left(E\left(\exp\left(\mathbf{t}.\mathbf{\xi}^{'}\right)\right)\right)=\\
\log\left(E\left(\exp\left(t_{1}\xi_{1}+\ldots+t_{l}\xi_{l}\right)\right)\right)=\\
\log\left(E\left(\exp\left(t_{1}\sum_{I_{1}}X_{j_{1}}+\ldots+t_{l}\sum_{I_{l}}X_{j_{l}}\right)\right)\right)=\\
\log\left(E\left(\exp\left(g_{1}\left(\mathbf{t}\right)X_{1}+\ldots+g_{J}\left(\mathbf{t}\right)X_{J}\right)\right)\right)=\\
\log\left(E\left(\exp\left(g\left(\mathbf{t}\right).\mathbf{X}^{'}\right)\right)\right)=K_{\mathbf{X}}\left(g\left(\mathbf{t}\right)\right)\label{eq:cum_gen_fun_suma}
\end{multline}

Function $g:\mathbb{R}^{l}\rightarrow\mathbb{R}^{J}$ is a vector
function defined by

\begin{eqnarray}
g\left(\mathbf{t}\right) & = & \left(g_{1}\left(\mathbf{t}\right),\ldots,g_{J}\left(\mathbf{t}\right)\right)\nonumber \\
g_{j}\left(\mathbf{t}\right) & = & \mathbf{t}.\left(\mathbf{1}\left(j\in I_{1}\right),\ldots,\mathbf{1}\left(j\in I_{l}\right)\right)^{'}\label{eq:transf_cums}
\end{eqnarray}
where
\[
\mathbf{1}\left(j\in I_{k}\right)=\begin{cases}
1, & j\in I_{k}\\
0, & j\notin I_{k}
\end{cases}
\]

It is hence possible to find the cumulant generating function of random
vector $\mathbf{\xi}\in\mathbb{R}^{l}$ in terms of that of the original
vector $\mathbf{X}\in\mathbb{R}^{J}$. If we know the c.g.f. of the
original random vector $\mathbf{X}$, then the cumulants, the cumulant
generating function, and hence the approximate joint density of the
aggregated variables, via Saddlepoint approximation at (\ref{eq:Saddlepoint})
of $\mathbf{\xi}\in\mathbb{R}^{l}$ are also determined (see section
\ref{sec:Illustration:-Runoff-to}). We can use this fact in order
to fit the modeol for $\mathbf{X}$ in such a way that the interesting
interaction manifestation (the sums of components) are explicitly
considered in the estimation.

\subsubsection{Joint probabilities of (multivariate) marginals}

Joint marginal distributions are usually important interaction manifestations.
Given a sub-vector $\mathbf{Y}:=\left(X_{j_{1}},\ldots,X_{j_{k}}\right)$
of $\mathbf{X}$, in order to find probabilities of the form 
\[
\Pr\left(X_{j_{1}}\geq x_{j_{1}},\ldots,X_{j_{k}}\geq x_{j_{k}}\right)
\]
one should in principle integrate expression (\ref{eq:Saddlepoint}),
for the c.f.g. of $\mathbf{Y}$.

In the uni-variate case, it is a well-established practice \citet{Huzurbazar_saddlepoint1999}
to employ instead an accurate approximation to that integral, which
is due to \citet{lugannani_rice1980}. Namely, in the univariate case,
we have: 
\begin{multline}
F_{X}\left(x_{0}\right)\approx\intop_{-\infty}^{x_{0}}\frac{\exp\left(K_{X}\left(\hat{\lambda}\left(x\right)\right)-x\hat{\lambda}\left(x\right)\right)}{\left(2\pi\right)^{1/2}\left(\frac{d^{2}K_{X}\left(\mathbf{\lambda}\right)}{d\lambda^{2}}\mid_{\lambda=\hat{\lambda}\left(x\right)}\right)^{1/2}}dx\\
\approx\Phi\left(r\right)+\phi\left(r\right)\left\{ \frac{1}{r}-\frac{1}{q}\right\} \label{eq:Lugganani}
\end{multline}

Where $\hat{\tau}$ is such that $K_{X}^{'}\left(\hat{\tau}\right)=x_{0}$,
and: 
\begin{eqnarray*}
r & = & sign\left(\hat{\tau}\right)\left\{ 2\left[\hat{\tau}x_{0}-K_{X}\left(\hat{\tau}\right)\right]\right\} ^{\frac{1}{2}}\\
q & = & \hat{\tau}\left\{ \frac{d^{2}K_{X}\left(\lambda\right)}{d\lambda^{2}}\mid_{\lambda=\hat{\tau}}\right\} ^{\frac{1}{2}}
\end{eqnarray*}

Thus, one must not perform the numerical integration at all. 

For the multivariate case, \citet{kolassa2010multivariate} have provided
a generalization of the Lugannani-Rice formula, which produces an
approximation to probability $\Pr\left(\mathbf{Y}\geq\mathbf{y}\right)$
of order $O\left(n^{-1}\right)$, for $\mathbf{X}\in\mathbb{R}^{J}$.
This formula is extremely complicated and writing it here will most
likely obscure rather than clarify anything. Only the probability
distribution function of a multivariate Normal distribution with covariance
matrix given by

\[
\Gamma_{ij}=\frac{\partial^{2}}{\partial t_{i}\partial t_{j}}K_{\mathbf{X}}\left(\mathbf{t}\right)\mid_{\mathbf{t}=\mathbf{0}}
\]
must be computed. For this task there are accurate methods available
for up to 20 dimensions \citet{Genz93comparisonof}.

If one intends to deal with vectors of dimension at most 5, corresponding
to multidimensional marginals of the random field modeled, we consider
more convenient to use numerical integration of (\ref{eq:Saddlepoint}).
For higher dimensions it would be better to use the result of \citet{kolassa2010multivariate}
in order to avoid difficult and inaccurate integrations.

\subsubsection{Differential entropy}

This also an important interaction manifestation, often encountered
in statistical research. Using the shorthand notation of \ref{eq:shorthand_not},
define $Z\left(\mathbf{x}\right):=\frac{1}{3!}\kappa^{j_{1},j_{2},j_{3}}h_{j_{1}j_{2}j_{3}}\left(\mathbf{x};\Gamma\right)$.
\citet{barros_2005_2005} studies an approximation to the differential
entropy of $\mathbf{X}$, which utilizes only the first correction
term in \ref{eq:edgeworth_series}:

\begin{multline}
\intop f_{\mathbf{X}}\left(\mathbf{x}\right)\log\left(f_{\mathbf{X}}\left(\mathbf{x}\right)\right)d\mathbf{x}=H\left(\phi_{\Gamma}\right)-\intop f_{\mathbf{X}}\left(\mathbf{x}\right)\log\left(\frac{f_{\mathbf{X}}\left(\mathbf{x}\right)}{\phi_{\Gamma}\left(\mathbf{x}\right)}\right)d\mathbf{x}\\
\approx H\left(\phi_{\Gamma}\right)-\int\phi_{\Gamma}\left(\mathbf{x}\right)\left(1+Z\left(\mathbf{x}\right)\right)\log\left(1+Z\left(\mathbf{x}\right)\right)d\mathbf{x}\\
\approx H\left(\phi_{\Gamma}\right)-\int\phi_{\Gamma}\left(\mathbf{x}\right)\left(Z\left(\mathbf{x}\right)+\frac{1}{2}Z\left(\mathbf{x}\right)^{2}\right)d\mathbf{x}=H\left(\phi_{\Gamma}\right)-\frac{1}{12}\Big\{\sum_{j=1}^{J}\left(k^{j,j,j}\right)^{2}\\
+3\sum_{i,j=1,i\neq j}^{J}\left(\kappa^{i,i,j}\right)^{2}+\frac{1}{6}\sum_{i,j,k=1,i<j<k}^{J}\left(\kappa^{i,j,k}\right)^{2}\Big\}\label{eq:Van_Hulle_approx}
\end{multline}

The value of $H\left(\phi_{\Gamma}\right)$ can be found in closed
form, $H\left(\phi_{\Gamma}\right)=\frac{1}{2}\log\left(\det\left(\Gamma\right)\right)+\frac{J}{2}\log\left(2\pi\right)+\frac{J}{2}$.
The approximation (\ref{eq:Van_Hulle_approx}) is accurate to order
$O\left(n^{-2}\right)$.

\subsection{Summarizing}

As we have seen in this section, joint cumulants provide us not only
with a lower bound for the number of variables interacting within
a vector; joint cumulants can also be connected with relevant interaction
manifestations, that may have a specific subject-matter interpretation.
The fitting of these interaction manifestations can be integrated
into model parameter estimation explicitly. 

Please note that the idea of the approach here presented is \emph{not}
to estimate the joint cumulants of a random vector by means of sample
joint cumulants. The goal is to fit as well as possible the relevant
interaction manifestation. The joint cumulants of the vector are fitted,
in that the set of joint cumulants that best recovers the interesting
interaction manifestation is kept, even if they are very different
from the sample ones.

\section{\label{sec:Illustration:-Runoff-to}Illustration: Extending the Gaussian
model }

In this section we illustrate the ideas put forward in this paper,
taking as interesting interaction manifestation the distribution of
the sums of sub-vectors of a random vector, mentioned in section \ref{sec:Interaction-parameters-versus-manifestations}. 

Another, more extended application of these ideas in the context of
spatial statistics can be found at \citet{ellipticalSpatialRodriguezBardossy}.
See also section \ref{sub:A-model-for-spatial statistics}.

The multivariate Normal model is a widely applied model in multivariate
analysis. A random vector $\mathbf{X}\in\mathbb{R}^{J}$ having mean
vector $\mathbf{m}$ and covariance matrix $\Gamma$, has c.g.f. given
by, 
\begin{equation}
K_{\mathbf{X}}\left(\mathbf{s}\right)=\mathbf{s}.\mathbf{m}^{T}+\frac{1}{2}\mathbf{s}\Gamma\mathbf{s}^{T}\label{eq:Gauss_cgf-1}
\end{equation}

A similar c.g.f. was studied by \citet{Steyn19931}, 
\begin{equation}
K_{\mathbf{X}}\left(\mathbf{s}\right)=\mathbf{s}.\mathbf{m}^{T}+\frac{c_{1}}{1!}\left(\frac{1}{2}\mathbf{s}\Gamma\mathbf{s}^{T}\right)+\frac{c_{2}}{2!}\left(\frac{1}{2}\mathbf{s}\Gamma\mathbf{s}^{T}\right)^{2}+\frac{c_{3}}{3!}\left(\frac{1}{2}\mathbf{s}\Gamma\mathbf{s}^{T}\right)^{3}+\ldots\label{eq:archetypal_cgf-1}
\end{equation}

Indeed, this c.g.f. reduces to that of the Gaussian model by setting
$c_{1}=1$ and $c_{r>1}=0$. In order to avoid identifiability problems
of the covariance matrix, we set $c_{1}=1$ and declare $\Gamma$
to be a true covariance matrix. This model is treated in detail at
\citet{ellipticalSpatialRodriguezBardossy}, in the context of spatial
statistics; it is shown at \citet{ellipticalSpatialRodriguezBardossy}
that it covers a span of tail dependence going from zero (i.e. Normal)
to that of the Student-t.

\subsection{Some data}

\begin{figure}
\begin{centering}
\includegraphics[width=0.75\textwidth]{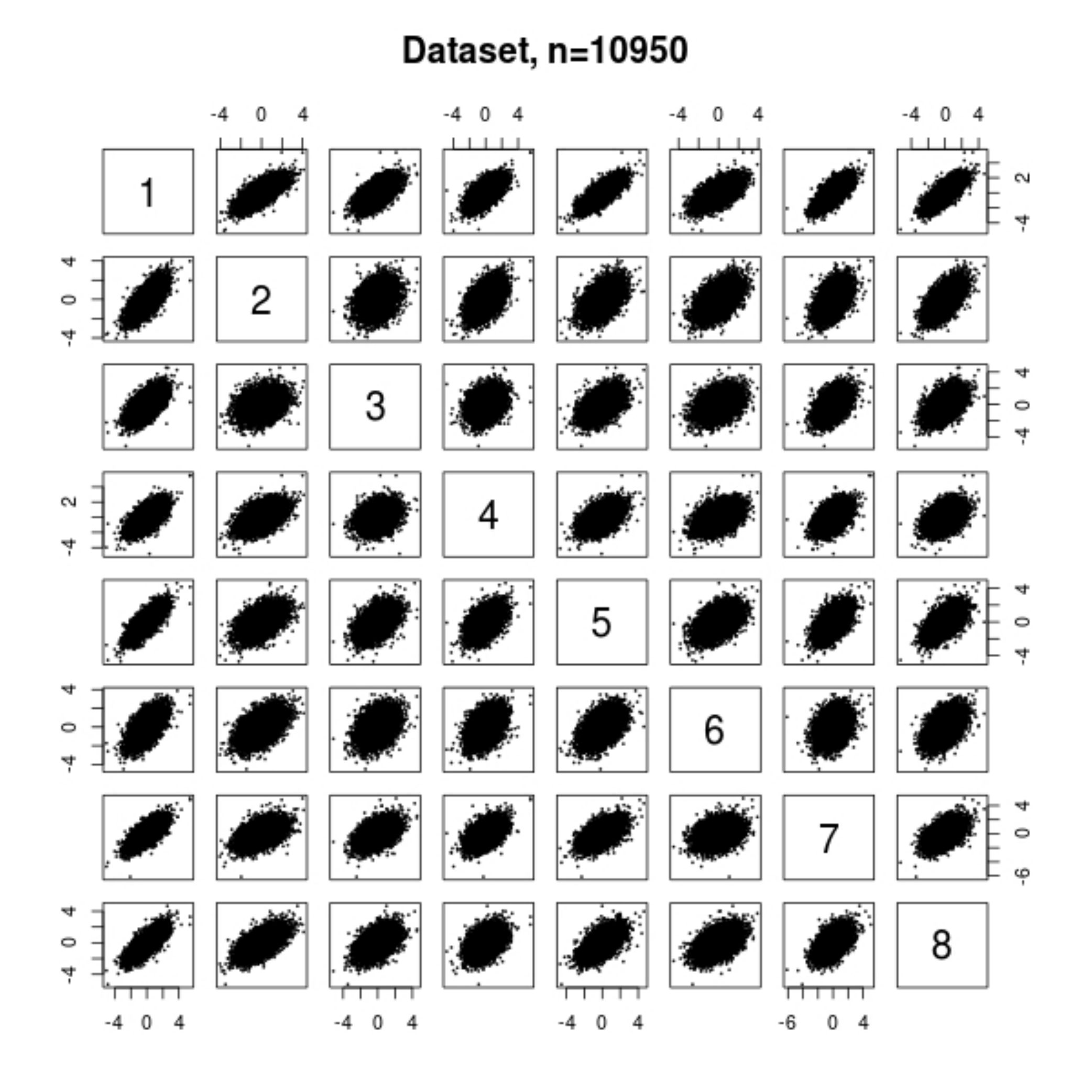}
\par\end{centering}

\caption{\label{fig:dataset_Y}8-dimensional test data set}

\end{figure}

In figure \ref{fig:dataset_Y} an 8-dimensional dataset is presented,
with a size of $n=10950$ realization. This dataset may represent
the daily (log) return of 8 stocks, or they could represent some daily
measured environmental variable at 8 locations, possibly after transformation.
In either case this dataset would amount to a 30 year record. A plot
of the data appears in figure \ref{fig:dataset_Y}. We are interested
in fitting a model that recovers properly the distribution of the
sum of the components of the 8-dimensional random vector, $S_{\mathbf{X}}=\sum_{i=1}^{8}X_{i}$. 

We shall employ the model given by c.g.f. (\ref{eq:archetypal_cgf-1}),
due to the shape of data, and to the flexibility of the mentioned
model to represent tail dependence. Specifically, we are interested
in fitting a model that captures the correlation among the 8 components
properly, but \emph{additionally} provides a good estimation to the
distribution of interaction manifestation 
\begin{equation}
S_{\mathbf{X}}=\sum_{j=1}^{8}X_{j}
\end{equation}

We assume for simplicity a mean vector $\mathbf{m}=\left(0,\ldots,0\right)$
of zeros (otherwise, data could be standardized to have zero means,
first). As in section \ref{sub:Connection-of-dependence-sumas}, we
have that the c.g.f. of $S_{\mathbf{X}}$ is given by
\begin{equation}
K_{S_{\mathbf{X}}}\left(t\right)=K_{\mathbf{X}}\left(g\left(t\right)\right)=\frac{c_{1}}{1!}\left(\frac{1}{2}g\left(t\right)\Gamma g\left(t\right)^{T}\right)+\frac{c_{2}}{2!}\left(\frac{1}{2}g\left(t\right)\Gamma g\left(t\right)^{T}\right)^{2}+\frac{c_{3}}{3!}\left(\frac{1}{2}g\left(t\right)\Gamma g\left(t\right)^{T}\right)^{3}+\ldots\label{eq:CGF_Suma}
\end{equation}
where 
\begin{equation}
g\left(t\right)=\underbrace{\left(t,\ldots,t\right)}_{8}
\end{equation}

\subsection{\label{sub:Parameter-estimation}Parameter estimation}

Our estimating strategy consists of:

\paragraph*{Step 1: Estimate Covariance matrix $\Gamma$}

In this way capture much of the 8-dimensional dependence structure.
Since our model is a member of the elliptical family, we can use the
estimator for the correlation matrix which uses Kendall's $\tau$
correlation coefficient (see \citet{muller_kendalls_2003}),
\begin{equation}
\hat{cor}\left(X_{i},X_{j}\right)=\sin\left(\frac{\pi}{2}\tau\left(X_{i},X_{j}\right)\right)
\end{equation}
whereby a complete correlation matrix, $\hat{R}$, is obtained.

Then the covariance matrix estimate can be found by 
\begin{equation}
\hat{\Gamma}=\Sigma^{\frac{1}{2}}\hat{R}\Sigma^{\frac{1}{2}}
\end{equation}
with 
\[
\Sigma=\left(\begin{array}{ccc}
S^{2}\left(X_{1}\right) & \ldots & 0\\
\vdots & \ddots & \vdots\\
0 & \ldots & S^{2}\left(X_{8}\right)
\end{array}\right)
\]
and $S^{2}\left(X_{j}\right)$ stands for the sample variance of $X_{j}$.
This procedure was followed, resulting in the covariance matrix given
in table \ref{tab:Estimated-covariance-for}, at the appendix.

Alternatively, if data represents an environmental variable sampled
at several locations, standard geostatistical tools can be used to
estimate $\Gamma$ (see \citet{ellipticalSpatialRodriguezBardossy}).
The covariance matrix will be in the following considered as known.

\paragraph*{Step 2: Interaction manifestation fitting}

We do this in a ``method-of-moments'' fashion (method of cumulants,
should we say). The $r$-th order cumulant of $S_{\mathbf{X}}$, $\kappa_{r}\left(S_{\mathbf{X}}\right)$,
can be found by differentiating (\ref{eq:CGF_Suma}) $r$ times with
respect to $t$, and then setting $t=0$. Performing the necessary
computations, one has for the mean and the variance: 
\begin{eqnarray}
\kappa_{1}\left(S_{\mathbf{X}}\right) & = & 0\\
\kappa_{2}\left(S_{\mathbf{X}}\right) & = & \frac{c_{1}}{1!}\frac{2}{2}\sum_{i,j=1}^{8}\Gamma_{ij}
\end{eqnarray}
and in general, odd-ordered cumulants will be zero, while even-ordered
cumulants are given by
\begin{equation}
\kappa_{2r}\left(S_{\mathbf{X}}\right)=\frac{c_{r}}{r!}\frac{\left(2r\right)!}{2^{r}}\left(\sum_{i_{1},\ldots,i_{r}=1}^{8}\sum_{j_{1},\ldots,j_{r}=1}^{8}\Gamma_{i_{1}j_{1}}\ldots\Gamma_{i_{r}j_{r}}\right)\label{eq:cumulantes_de_suma}
\end{equation}

We compute the sample cumulants, $\hat{\kappa}_{2r}$ (for $r=1,2,3$),
of $S_{\mathbf{X}}$. These are found to be 37.426, 463.509 and 105098.112,
respectively. Substituting these sample cumulants for the theoretical
cumulants in (\ref{eq:cumulantes_de_suma}), and using the already
available covariance matrix, $\Gamma$, we can estimate $c_{1}$,
$c_{2}$ and $c_{3}$. These estimates are given by $\hat{c}_{1}=0.999$,
$\hat{c}_{2}=0.1101$ and $\hat{c}_{3}=0.1332$. Note that by considering
cumulants of $S_{\mathbf{X}}$ of order $\geq4$, we can capture important
tail characteristics of its distribution.

\subsection{Evaluation of the fit}

We use the Monte Carlo approach to evaluate the fit carried out in
the previous sub-section. One can sample from a random vector, $\mathbf{Y}\in\mathbb{R}^{8}$,
having c.g.f. as in (\ref{eq:archetypal_cgf-1}), by sampling two
independent random variables: 1. a non-negative random variable $V>0$,
with cumulants $c_{1},\ldots,c_{r}$ (in our case, $r=3)$; 2. a normally
distributed random vector $\mathbf{X}\sim N\left(\mathbf{0},\Gamma\right)$.
Then one sets: 
\begin{equation}
\mathbf{Y}=\mathbf{m}+\sqrt{V}\times\mathbf{X}\label{eq:construccion_Kano}
\end{equation}

For more details, the reader is referred to \citet{ellipticalSpatialRodriguezBardossy}. 

\begin{figure}
\begin{centering}
\includegraphics[width=0.75\textwidth]{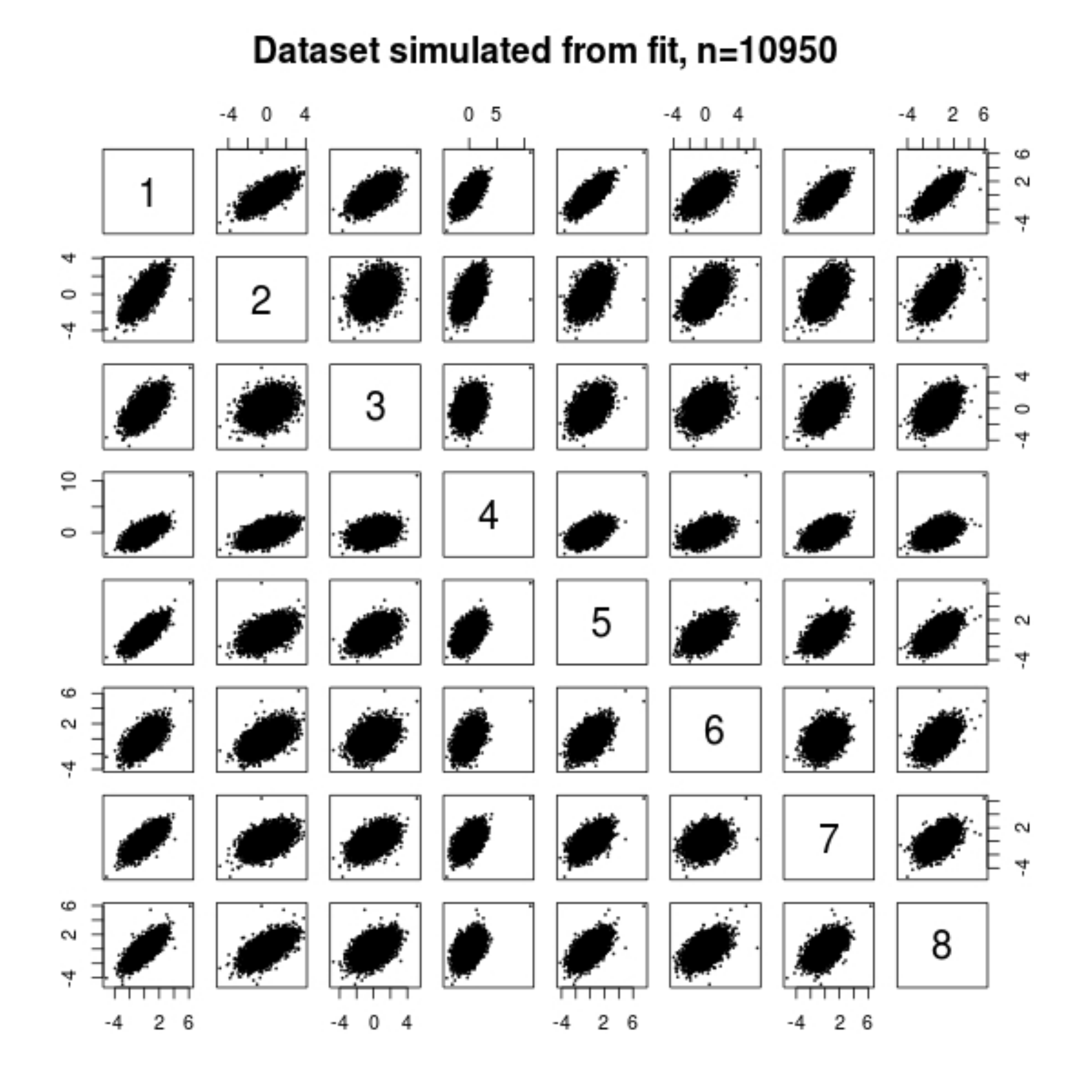}
\par\end{centering}

\caption{\label{fig:One-sample-of-Y-new}One sample of size n=10950, generated
using the parameters fitted in section \ref{sub:Parameter-estimation}.}

\end{figure}

We fitted $V$ as a mixture of 5 gamma random variables, in such a
way that the cumulants of this mixture are $\hat{c}_{1}=0.999$, $\hat{c}_{2}=0.1101$
and $\hat{c}_{3}=0.1332$, up to a small error. Then we were able
to simulate 1000 samples of $\mathbf{Y}$, each of size $n=10950$,
using the fitted parameters. One of the realizations is shown in figure
\ref{fig:One-sample-of-Y-new}. Note that the covariance structure
is mostly recovered, though there are some outliers of a magnitude
somewhat larger than those displayed in figure \ref{fig:dataset_Y}.
This is because, once we fitted covariance matrix $\Gamma$, we focus
on recovering the distribution of the sum of the components of the
vector $\mathbf{X}$, i.e. $S_{\mathbf{X}}$. The outliers there presented
are part of the mechanism that helps recover the distribution of the
components sum.

To see how well the fitted parameters reproduce $S_{\mathbf{X}}$,
we present several sample quantiles of it, together with confidence
bands built out of the 1000 Monte Carlo simulations. See table \ref{tab:Representative-quantiles-of}.
We see an excellent cover of the given quantiles, particularly at
the tails of the distribution of $S_{\mathbf{X}}$. 

\begin{table}
\begin{centering}
\begin{tabular}{|c||c||c||c|}
\hline 
Quantile (\%) & 2.75\% & 97.5\% & Observed\tabularnewline
\hline 
\hline 
0 (min) & -41.921 & -22.644 & -29.191\tabularnewline
\hline 
\hline 
0.1 & -21.549 & -18.876 & -20.72\tabularnewline
\hline 
\hline 
0.5 & -16.956 & -15.72 & -17.032\tabularnewline
\hline 
\hline 
1 & -15.033 & -14.124 & -14.771\tabularnewline
\hline 
\hline 
5 & -10.248 & -9.748 & -10.049\tabularnewline
\hline 
\hline 
10 & -7.897 & -7.518 & -7.774\tabularnewline
\hline 
\hline 
20 & -5.159 & -4.838 & -5.194\tabularnewline
\hline 
\hline 
25 & -4.138 & -3.838 & -4.212\tabularnewline
\hline 
\hline 
50 & -0.136 & 0.137 & -0.18\tabularnewline
\hline 
\hline 
75 & 3.836 & 4.145 & 4.013\tabularnewline
\hline 
\hline 
80 & 4.84 & 5.155 & 4.987\tabularnewline
\hline 
\hline 
90 & 7.507 & 7.903 & 7.573\tabularnewline
\hline 
\hline 
95 & 9.765 & 10.273 & 9.911\tabularnewline
\hline 
\hline 
99 & 14.114 & 15.058 & 14.293\tabularnewline
\hline 
\hline 
99.5 & 15.718 & 17.034 & 16.01\tabularnewline
\hline 
\hline 
99.9 & 18.93 & 21.625 & 20.159\tabularnewline
\hline 
\hline 
99.99 & 21.908 & 29.07 & 28.542\tabularnewline
\hline 
\hline 
100 (max) & 22.897 & 43.735 & 28.983\tabularnewline
\hline 
\end{tabular}
\par\end{centering}

\caption{\label{tab:Representative-quantiles-of}Representative quantiles of
$S_{\mathbf{X}}$ and confidence bands of 1000 Monte Carlo simulations
of 10950 sized samples each. The parameters fitted in section \ref{sub:Parameter-estimation}
have been used for the simulation. Simulations reproduce quantiles
very similar to those observed.}

\end{table}

\begin{figure}
\begin{centering}
\includegraphics[width=0.75\textwidth]{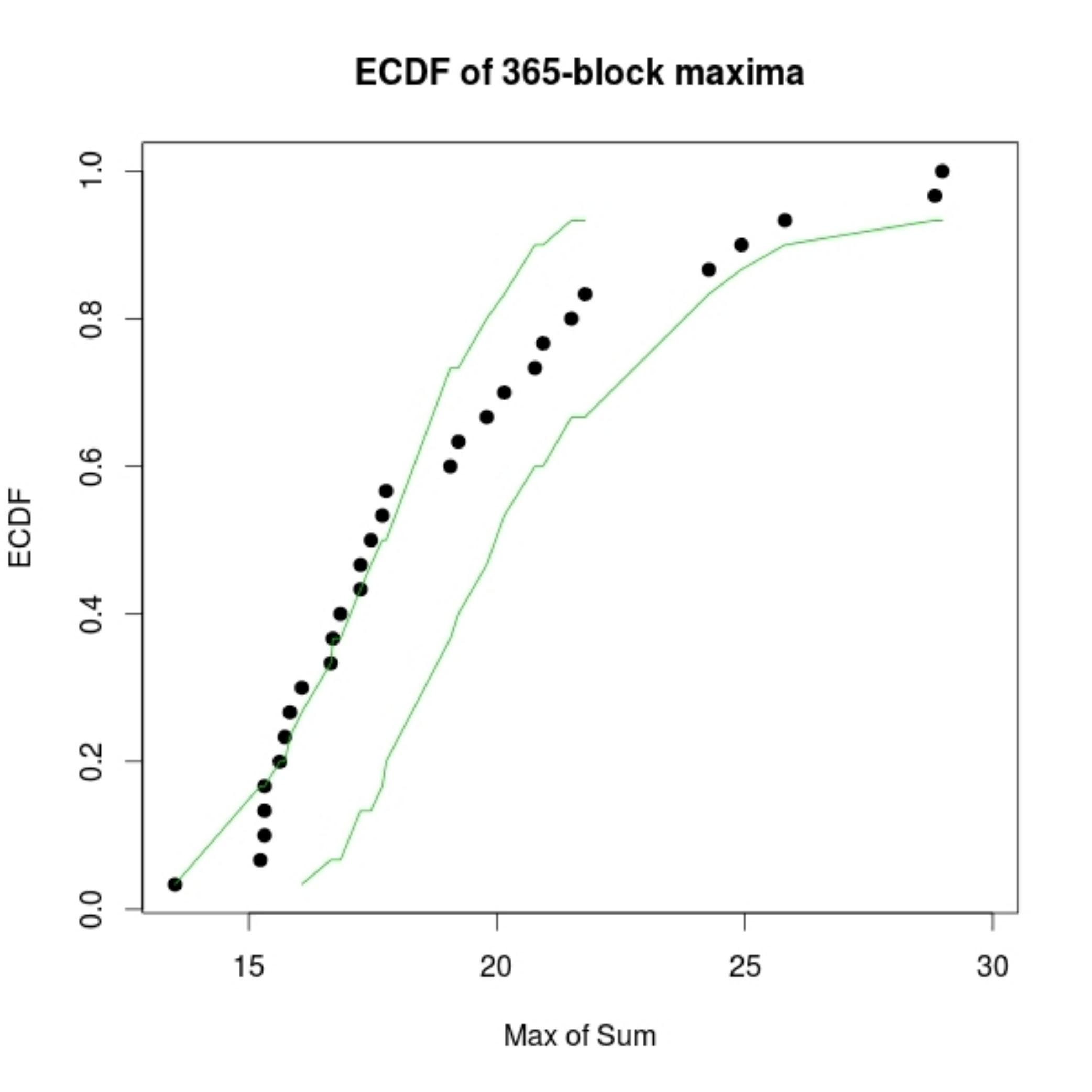}
\par\end{centering}

\caption{\label{fig:Empirical-Cumulative-Distributio-maxima}Empirical Cumulative
Distribution Function of the 365-block maxima, made out of the 10950
sized sample presented at figure \ref{fig:dataset_Y}. Monte Carlo
simulation based 95\% confidence bands have been added from data simulated
using the parameters fitted in this section.}

\end{figure}

Additionally, the distribution of the 365-block maxima of the components
sums is also acceptably recovered. In figure \ref{fig:Empirical-Cumulative-Distributio-maxima}
we show the empirical distribution function of the 30 sample 365-block
maxima (i.e. yearly maxima). The Monte Carlo based 95\% confidence
bands for the 365-block maxima of $S_{\mathbf{X}}$ are also presented
in figure \ref{fig:Empirical-Cumulative-Distributio-maxima}.

\subsection{More complicated questions}

The techniques presented in this section can also be used to investigate
more complex situations. For example, one would like to model jointly
the random variables
\begin{eqnarray}
Z_{1} & := & X_{1}+\ldots+X_{4}\\
Z_{2} & := & X_{5}+\ldots+X_{8}
\end{eqnarray}

This may be the case if each group of components, $X_{1},\ldots,X_{4}$
and $X_{5},\ldots,X_{8}$, refers each to a geographical area (in
environmental modeling); or if there is some economical reason to
group them (stock price modeling). We may then wish to model the distributions
of $Z_{1}$ and $Z_{2}$, but also model properly at least the correlation
between them.

Applying a similar computation as before, we find that
\begin{equation}
cov\left(Z_{1},Z_{2}\right)=\frac{c_{1}}{2}\sum_{i=1}^{4}\sum_{j=5}^{8}\Gamma_{ij}\label{eq:cov_ZZ}
\end{equation}
for the covariance. Regarding each $Z_{j}$, all odd-ordered cumulants
are zero, whereas all even-ordered cumulants are given by
\begin{equation}
\kappa_{2r}\left(Z_{j}\right)=\left(2r-1\right)!\times c_{r}\times\left(\frac{R_{j}}{2}\right)^{r}\label{eq:cums_ZZ}
\end{equation}
for $j=1,2$, where
\begin{eqnarray*}
R_{1} & = & 2\sum_{i=1}^{4}\Gamma_{ii}+4\sum_{1<i<j<4}\Gamma_{ij}\\
R_{2} & = & 2\sum_{i=5}^{8}\Gamma_{ii}+4\sum_{5<i<j<8}\Gamma_{ij}
\end{eqnarray*}

Using equations (\ref{eq:cov_ZZ}) and (\ref{eq:cums_ZZ}), and the
sample estimates for these quantities, we can fit parameters $c_{1},\ldots,c_{r}$
of (\ref{eq:archetypal_cgf-1}), as in section \ref{sub:Parameter-estimation}.
In this new case, we shall have parameters that reproduce well the
correlation among the aggregation vectors, and produce a good match
of the cumulants of each marginal distribution, thereby modeling each
marginal adequately.

\subsection{\label{sub:A-model-for-spatial statistics}A model for Spatial Statistics}

The model given by (\ref{eq:archetypal_cgf-1}) can be used to incorporate
multivariate interdependence into a spatial model, while keeping the
spatial consistency requirement that any subvector of the spatial
field must have the same distribution of the vector containing it
(\citet{ellipticalSpatialRodriguezBardossy}). The covariance matrix,
$\Gamma$, is thereby estimated using the standard technique of fitting
a covariance function to the spatially labeled data. The additional
parameters, $c_{2},c_{3},\ldots$ can be used to obtain a better fit
of any subject-matter relevant interaction manifestation.

Using data from the Saalach river catchment, in southeast Germany,
\citet{ellipticalSpatialRodriguezBardossy} fitted a spatio-temporal
model to precipitation data of nine gauging stations lying in the
catchment area. A model very similar to the one investigated by \citet{sanso_venezuelan_1999}
was fitted, because it can easily accommodate missing data as well
as the truncated nature of daily precipitation. The model relies on
a latent Gaussian field for spatial dependence modeling; that is a
latent model with cumulant generating function as (\ref{eq:archetypal_cgf-1}),
with $0=c_{2}=c_{3}=...$. 

\citet{ellipticalSpatialRodriguezBardossy} then study the implications
of selecting $\left(c_{1},\ldots,c_{5}\right)=\left(0.999,0.079,0.152,0.521,1.971\right)$,
instead of $c_{r>1}=0$ as in the original model by \citet{sanso_venezuelan_1999}.
As shown by \citet{ellipticalSpatialRodriguezBardossy}, a random
field with $\left(c_{1},\ldots,c_{5}\right)$ as above is practically
indistinguishable in its one and two dimensional marginal distributions
from a Gaussian field with the same covariance function and mean.
However, implications for the interaction manifestation ``average
of fields components'', where each component represents daily precipitation
over an 500 mt $\times$ 500 mt squared area on the Saalach river
catchment, are significant. 

\begin{figure}
\begin{centering}
\includegraphics[width=0.5\textwidth]{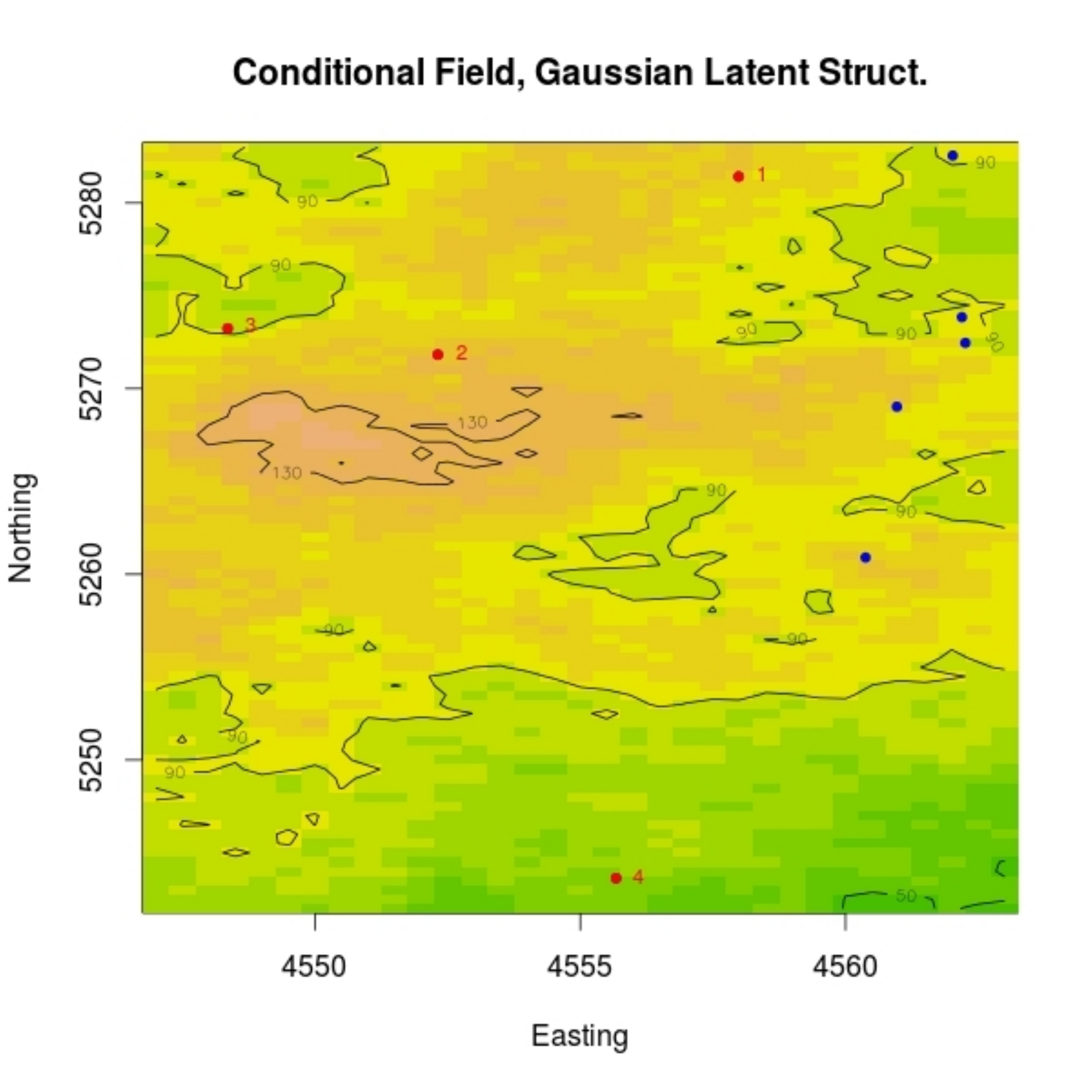}\includegraphics[width=0.5\textwidth]{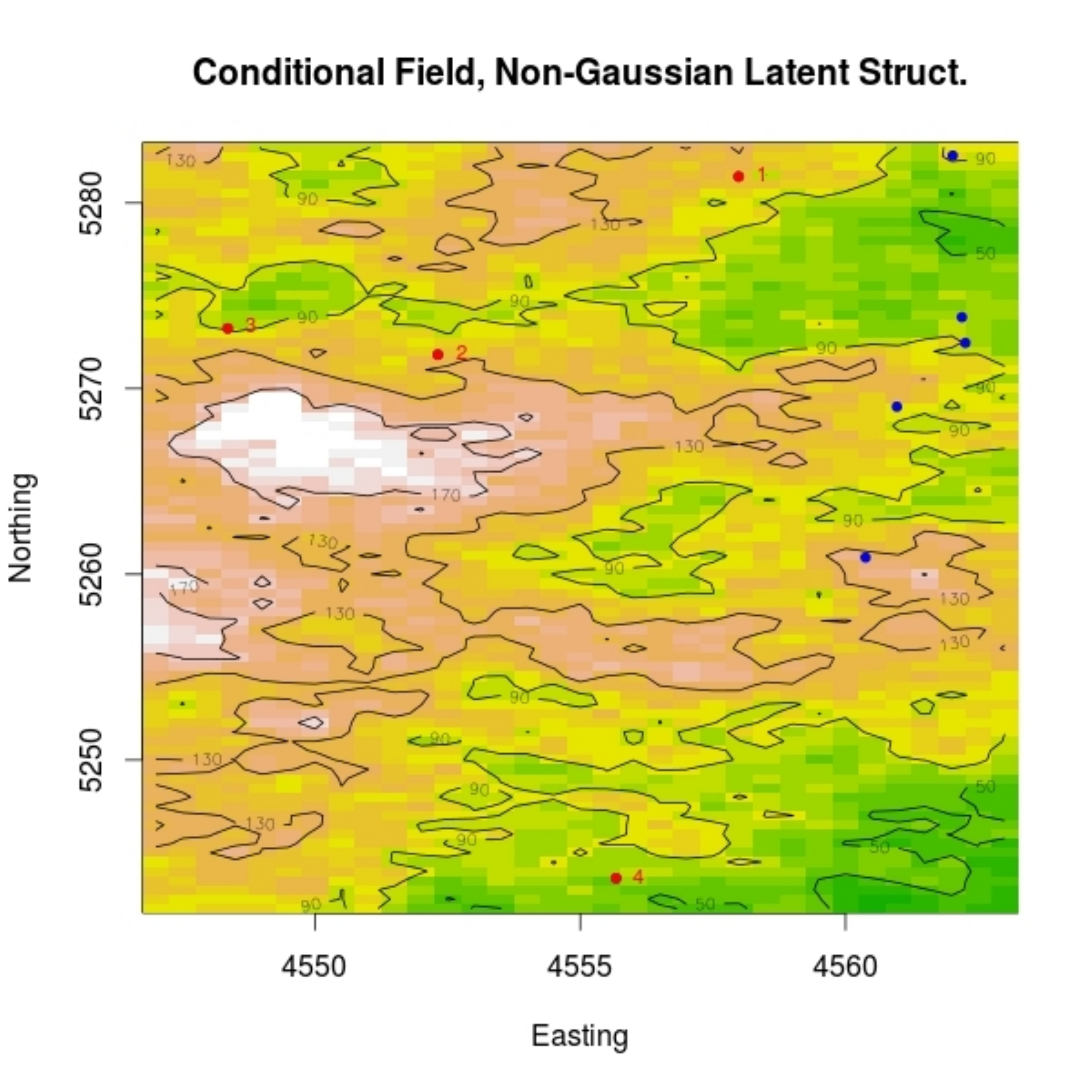}
\par\end{centering}

\caption{\label{fig:Two-conditionally-simulated-fields}Two conditionally simulated
fields for June 1st 2013, for part of the Saalach river catchment: Field with Gaussian latent structure (left),
and field with non-Gaussian latent structure (right). Stations providing
the observed data are indicated in red. Stations indicated by blue
points have no available data for that day. Note the intense precipitation
clusters predictable by the model with latent field having multivariate
interactions.}

\end{figure}

\begin{figure}
\begin{centering}
\includegraphics[width=0.75\textwidth]{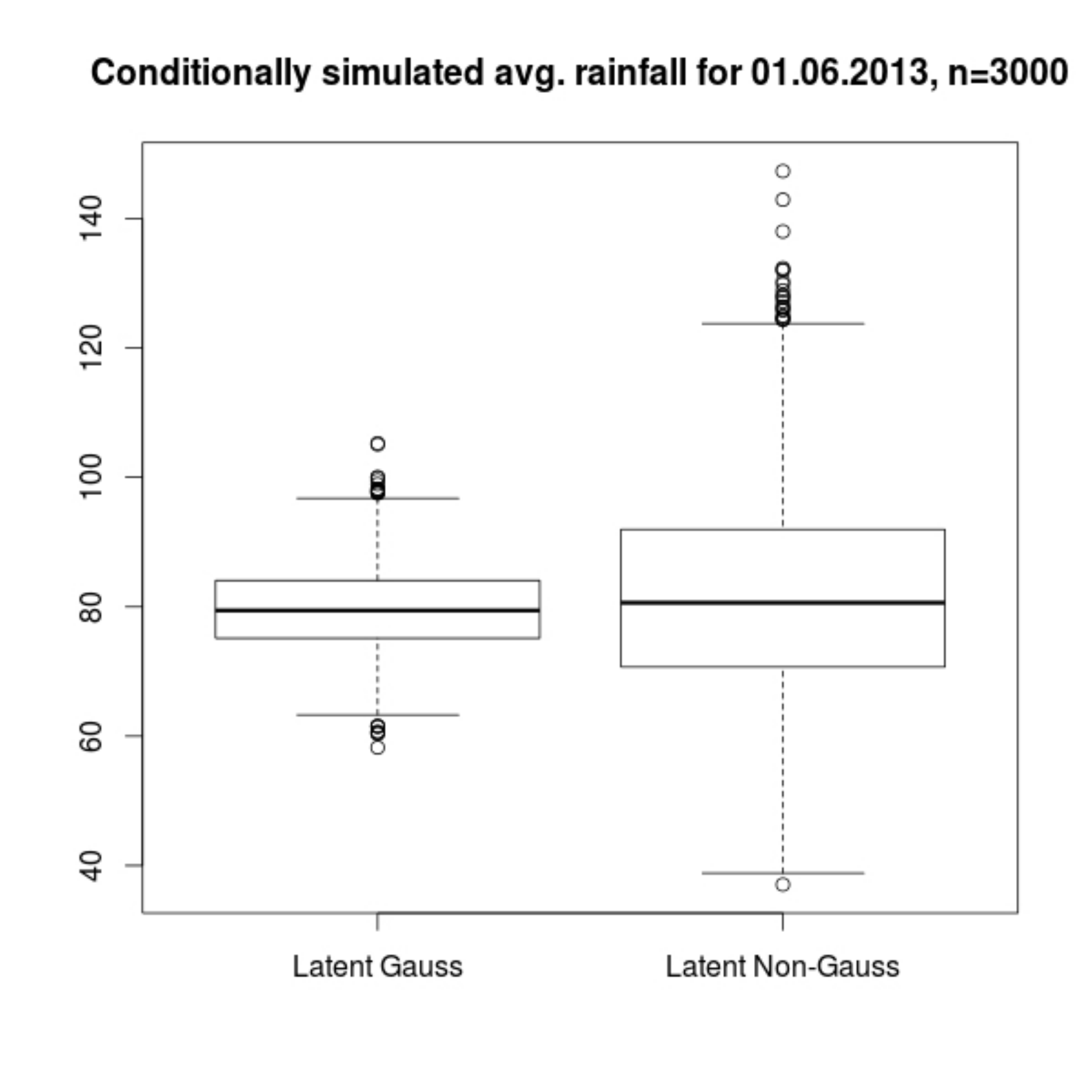}
\par\end{centering}

\caption{\label{fig:Boxplots-of-the-conditional}Boxplots of the average of
the conditionally simulated random fields for June 1st 2013, in millimeters,
for the Saalach river catchment. The field with high oder interacting
latent structure shows much more variability. In particular, average
precipitation over the catchment above 120 mm are quite probable under
this model.}

\end{figure}

The authors obtained 3000 conditional simulations, given the rainfall
data available, of the rainfall field over the Saalach river catchment
for June 1st 2013, a day of intense rainfall during the 2013 central
European floods. In figure \ref{fig:Two-conditionally-simulated-fields},
two of the obtained conditional fields are presented, using the Gaussian
and the almost-Gaussian latent structure. In figure \ref{fig:Boxplots-of-the-conditional},
we show the distribution of the conditional values of mean precipitation
over the catchment, for both latent structures. Note that the multivariate
interactions, hardly noticeable on the one and two dimensional marginal
distributions, increase dramatically the probability of a very high
mean precipitation over the studied catchment. The consequence is
that substantial under-estimation of flood return periods may me incurred,
if one does not account for interaction among more than tow components,
in one's spatio-temporal precipitation models.

\section{\label{sec:Discussion}Discussion}

An approach for considering interactions that go beyond correlations
has been presented. We have seen that the discrimination between interactions
``parameters'' and interactions ``manifestations'' can help to
circumvent two major problems one is confronted with when attempting
to quantify and model higher order interactions: the problem of interpretability,
by working with subject-matter relevant manifestations of interdependence;
and the problem of high dimensionality, by recoursing to joint cumulants
as building blocks of a dependence model. By using the cumulant generating
function, we are recoursing to a well-studied object: the characteristic
function of a distribution. 

As dimension of vector $\mathbf{X}$ increases, interactions of high
order may be more and more difficult to assess. For example, a random
vector having c.f.g. (\ref{eq:archetypal_cgf-1}), with $c_{1}=1$
, $c_{r}\approx0$ for $2\leq r\leq3$ but then $c_{r\geq4}\neq0$,
would have one and two dimensional marginals practically equal to
those of a Guassian distribution. But the interaction coefficients
of groups of 14 components or more will be very different, producing
very different interaction manifestations. The difference in the overall
dependence structures may grow tremendously as the dimension of the
random vector $\mathbf{X}$ grow (i.e. $J>>2$), even though these
fact may go totally unnoticed in the one and two dimensional marginal
analysis of data. 

In \citet{ellipticalSpatialRodriguezBardossy}, these issues are dealt
with and illustrated in the context of Spatial Statistics, where the
issue of low dimensionality is essential, and where interaction manifestations
can differ drastically between two models having very similar 1 and
2 dimensional marginals, due to the big dimension of the field.

\subsection*{Acknowledgments}

This research forms part of the Ph.D thesis of the first author, which
was funded by a scholarship of the German Academic Exchange Service
(DAAD). This Ph.D work was carried out within the framework of the
ENWAT program at the University of Stuttgart.

\appendix

\section{Review of Edgeworth Expansion and the Saddlepoint Approximation}

We recall well-known results about density approximation. Details
for all topics of this appendix can be found in \citet{barndorff-nielsen_asymptotic_1990,kolassa2006series};
we present here just the approximations, in the context of a distribution
having a probability density function. 

The \emph{Edgeworth Expansion} is a series expansion of the probability
density and of the probability distribution in terms of the joint
cumulants (performing as coefficients) and of the multivariate normal
distribution (performing as basis function). 

We employ below the shorthand notation for summations used in \citet{barndorff-nielsen_asymptotic_1990},
in order to avoid an overflow of symbols in these pages. Arrays are
represented by symbols with superscripts and under-scripts. For example
a matrix is represented by $a^{i,j}$ or by $b_{ij}$. An array with
three dimensions would be $c^{i,j,k}$ or $d_{ijk}$, and so on. The
product of these symbols indicates summation along all dimensions
for which the index is repeated. For example the term $\frac{1}{6\sqrt{n}}\kappa^{j_{1},j_{2},j_{3}}h_{j_{1}j_{2}j_{3}}$,
to be used below, should be interpreted as 
\begin{equation}
\frac{1}{6\sqrt{n}}\kappa^{j_{1},j_{2},j_{3}}h_{j_{1}j_{2}j_{3}}=\frac{1}{6\sqrt{n}}\sum_{j_{1}=1}^{J_{1}}\sum_{j_{2}=1}^{J_{2}}\sum_{j_{3}=1}^{J_{3}}\kappa^{j_{1},j_{2},j_{3}}h_{j_{1}j_{2}j_{3}}\label{eq:shorthand_not}
\end{equation}
where, for example,
\begin{eqnarray*}
\kappa^{j_{1},j_{2},j_{3}} & = & cum\left(X_{j_{1}},X_{j_{2}},X_{j_{3}}\right)\\
\kappa^{j_{1},j_{2},j_{3},j_{4}} & = & cum\left(X_{j_{1}},X_{j_{2}},X_{j_{3}},X_{j_{4}}\right)
\end{eqnarray*}

Let $\text{\ensuremath{\mathbf{Z}}}\in\mathbb{R}^{J}$ be a random
vector with probability density function $f$. Assume also, without
loss of generality, that $\mathbf{Z}$ has mean a vector of zeros,
a $J\times J$ covariance matrix $\kappa^{i,j}=\Gamma$, and joint
cumulants $\text{\ensuremath{\left\{  \kappa^{j_{1},j_{2},j_{3}}\right\} } ,\text{\ensuremath{\left\{  \kappa^{j_{1},j_{2},j_{3},j_{4}}\right\} } }},\ldots$.
If we have a random sample of $n$ i.i.d. random vectors with the
same distribution as $\mathbf{Z}$, namely $\mathbf{Z}_{1},\ldots,\mathbf{Z}_{n}$,
then we can form the average random vector $\mathbf{X}=\frac{1}{n}\sum_{i=1}^{n}\mathbf{Z}_{i}$.
This latter random vector has a density function $f_{\mathbf{X}}$
which can be formally written as the following series expansion, in
terms of the summation shorthand notation:
\begin{multline}
f_{\mathbf{X}}\left(\mathbf{x}\right)=\phi_{\Gamma}\left(\mathbf{x}\right)\Big\{1+\frac{1}{6\sqrt{n}}\kappa^{j_{1},j_{2},j_{3}}h_{j_{1}j_{2}j_{3}}\left(\mathbf{x};\Gamma\right)+\frac{1}{24n}\kappa^{j_{1},j_{2},j_{3},j_{4}}h_{j_{1}j_{2}j_{3}j_{4}}\\
+\frac{1}{72n}\kappa^{j_{1},j_{2},j_{3}}\kappa^{j_{4},j_{5},j_{6}}h_{j_{1}j_{2}j_{3}j_{4}j_{5}j_{6}}\left(\mathbf{x};\Gamma\right)\Big\}+O\left(n^{-\frac{3}{2}}\right)\label{eq:edgeworth_series}
\end{multline}

Where $\phi_{\Gamma}$ is the multivariate Normal density function
with zero mean and covariance matrix $\Gamma$, and $h_{j_{1}\ldots j_{k}}\left(\mathbf{x};\Gamma\right)$
represents the evaluation at $\mathbf{x}$ of the k-order Hermite
polynomial determined by the identity 
\begin{equation}
\text{\ensuremath{\phi}}_{\Gamma}\left(\mathbf{x}\right)h_{j_{1}\ldots j_{k}}\left(\mathbf{x};\Gamma\right)=\left(-1\right)^{k}\frac{\partial^{k}\phi_{\Gamma}\left(\mathbf{x}\right)}{\partial x_{j_{1}}\ldots\partial x_{j_{k}}}\label{eq:Multi_Hermite}
\end{equation}

Actually, $\phi_{\Gamma}\left(\mathbf{x}\right)$ is a Normal approximation
to $f_{\mathbf{X}}$, and the factors within brackets are often referred
to as \textquotedbl{}correction terms\textquotedbl{}.

It could be protested that we have considered only the case of an
average $\mathbf{X}=\frac{1}{n}\sum_{i=1}^{n}\mathbf{Z}_{i}$ of random
vectors. However, if the distribution of $\mathbf{Z}$ is unimodal
and not wildly skewed or leptokurtic, then the Edgeworth Approximation
given in \ref{eq:edgeworth_series} is often a good approximation
in practice even with $n=1$, as we shall use it. After all, a random
variable does not have to be the result of averaging $n$ variables
in order to have cumulants as such an average variable. This is the
case of the chi-squared distribution with $n$ degrees of freedom,
for example, which can be interpreted as the sum of $n$ standard
Normal variables after raising each to the second power. 

The usefulness of retaining the dependence on $n$ is that we are
reminded of when the Edgeworth Expansion is useful in practice: When
the cumulants of $\mathbf{X}$, of which the density must be approximated,
do not explode as their order increases, i.e. they behave as if $\mathbf{X}$
were approximately an average.

The Edgeworth expansion is practically accurate near the expected
value of the distribution, but degenerates as one moves towards the
tails of the distribution.

The \emph{Saddlepoint Approximation}, also called ``tilted'' Edgeworth
Approximation, is a more accurate approximation to the density of
$\mathbf{X}$ at the tails, which we can apply if we know its cumulant
generating function $K_{\mathbf{X}}\left(\mathbf{t}\right)$. In the
context of considering $\mathbf{X}$ as the mean of $n$ copies of
$\mathbf{Z}$, the relation between the cumulant generating functions
is $K_{\mathbf{X}}\left(\mathbf{t}\right)=nK_{\mathbf{Z}}\text{\ensuremath{\left(\frac{\mathbf{t}}{\sqrt{n}}\right)}}$.
As mentioned above, we shall be using this approximations as if we
were dealing with a variable being the average of $n=1$ random variables.
Thus we remove in the following the dependence on such an underlying
$n$ and work directly with $K_{\mathbf{X}}\left(\mathbf{t}\right)$.

In order to introduce the Saddlepoint Approximation, assume for a
moment we are trying to find the Edgeworth Expansion not of $f_{\mathbf{X}}\left(\mathbf{x}\right)$,
but of a related family of density functions, defined in terms of
an auxiliary vector $\mathbf{\lambda}\in\mathbb{R}^{J}$,
\begin{equation}
f_{\mathbf{X}}\left(\mathbf{x};\mathbf{\lambda}\right)=\exp\left(\mathbf{x^{T}}.\mathbf{\lambda}-K_{\mathbf{X}}\left(\mathbf{\lambda}\right)\right)f_{\mathbf{X}}\left(\mathbf{x}\right)\label{eq:tilted}
\end{equation}

The idea is, for \emph{each} $\mathbf{x}\in\mathbb{R}^{J}$ to choose
the most advantageous value $\mathbf{\hat{\mathbf{\lambda}}}$ of
$\mathbf{\lambda}\in\mathbb{R}^{J}$ in order to make the Edgeworth
approximation $\hat{f}_{\mathbf{X}}\left(\mathbf{x};\mathbf{\lambda}\right)$
to $f_{\mathbf{X}}\left(\mathbf{x};\mathbf{\lambda}\right)$ as accurate
as possible. Of course, this will provide automatically an approximation
for $f_{\mathbf{X}}$, 
\[
\hat{f}_{\mathbf{X}}\left(\mathbf{x}\right)=\exp\left(K_{\mathbf{X}}\left(\mathbf{\hat{\lambda}}\right)-\mathbf{x^{T}}.\mathbf{\hat{\lambda}}\right)\hat{f}_{\mathbf{X}}\left(\mathbf{x};\mathbf{\hat{\lambda}}\right)
\]
which is in fact what we want. 

The optimum value $\mathbf{\hat{\lambda}}$ can be proved to be the
one fulfilling $\mathbf{x}=\nabla K_{\mathbf{X}}\text{\ensuremath{\left(\mathbf{\hat{\lambda}}\right)}}$,
for the particular $\mathbf{x}\in\mathbb{R}^{J}$ in question, because
then density $f_{\mathbf{X}}\left(\mathbf{x};\mathbf{\hat{\lambda}}\right)$
corresponds to a random vector having its mean at $\mathbf{x}$, where
the Edgeworth Approximation is most accurate. Now, under suitable
regularity conditions, the leading term of the Edgeworth expansion
of $f_{\mathbf{X}}\left(\mathbf{x};\mathbf{\hat{\lambda}}\right)$
is a multivariate Normal density with covariance matrix with entries
\[
\left(\hat{\Sigma}_{i,j}\right)=\frac{\partial^{2}K_{\mathbf{X}}\left(\mathbf{\lambda}\right)}{\partial\lambda_{i}\partial\lambda_{j}}\mid_{\mathbf{\lambda}=\mathbf{\hat{\lambda}}}
\]
evaluated at its mean; that is,

\[
f_{\mathbf{X}}\left(\mathbf{x};\mathbf{\hat{\lambda}}\right)\approx\frac{e^{0}}{\left(2\pi\right)^{J/2}\det\left(\Sigma\right)^{1/2}}
\]
Thus, the looked for approximation is given by
\begin{equation}
f_{\mathbf{X}}\left(\mathbf{x}\right)=\exp\left(K_{\mathbf{X}}\left(\mathbf{\hat{\lambda}}\right)-\mathbf{x^{T}}.\mathbf{\hat{\lambda}}\right)f_{\mathbf{X}}\left(\mathbf{x};\mathbf{\hat{\lambda}}\right)\approx\frac{\exp\left(K_{\mathbf{X}}\left(\mathbf{\hat{\lambda}}\right)-\mathbf{x^{T}}.\mathbf{\hat{\lambda}}\right)}{\left(2\pi\right)^{J/2}\det\left(\hat{\Sigma}\right)^{1/2}}\label{eq:Saddlepoint}
\end{equation}

The error of this approximation is of order $O\left(n^{-1}\right)$
for all $\mathbf{x}\in\mathbb{R}^{J}$, if the joint cumulants of
random vector $\mathbf{X}$ behave like an average of \emph{n} iid
random vectors. Suitable normalization can bring this order down to
$O\left(n^{-2}\right)$. 

In spite of the apparent disadvantage of having to re-compute the
density estimation for each $\mathbf{x}$, the computational cost
becomes considerably smaller than that of the Edgeworth Approximation
as dimension increases, since the number of multivariate Hermite polynomials
at \ref{eq:edgeworth_series} to evaluate increases exponentially
with the dimension of $\mathbf{x}$.

\section{Estimated Covariance for the illustration}

\begin{table}[H]
\begin{centering}
\begin{tabular}{|l||l||l||l||l||l||l||l||l|}
\hline 
 & 1 & 2 & 3 & 4 & 5 & 6 & 7 & 8\tabularnewline
\hline 
\hline 
1 & 1.003 & 0.716 & 0.624 & 0.638 & 0.767 & 0.616 & 0.714 & 0.768\tabularnewline
\hline 
\hline 
2 & 0.716 & 0.988 & 0.311 & 0.507 & 0.491 & 0.53 & 0.468 & 0.635\tabularnewline
\hline 
\hline 
3 & 0.624 & 0.311 & 1.009 & 0.291 & 0.47 & 0.369 & 0.496 & 0.488\tabularnewline
\hline 
\hline 
4 & 0.638 & 0.507 & 0.291 & 0.979 & 0.504 & 0.419 & 0.499 & 0.422\tabularnewline
\hline 
\hline 
5 & 0.767 & 0.491 & 0.47 & 0.504 & 0.991 & 0.486 & 0.55 & 0.599\tabularnewline
\hline 
\hline 
6 & 0.616 & 0.53 & 0.369 & 0.419 & 0.486 & 0.992 & 0.282 & 0.486\tabularnewline
\hline 
\hline 
7 & 0.714 & 0.468 & 0.496 & 0.499 & 0.55 & 0.282 & 1.024 & 0.52\tabularnewline
\hline 
\hline 
8 & 0.768 & 0.635 & 0.488 & 0.422 & 0.599 & 0.486 & 0.52 & 1.007\tabularnewline
\hline 
\end{tabular}
\par\end{centering}

\caption{\label{tab:Estimated-covariance-for}Estimated covariance for the
illustrative dataset of section \ref{sec:Illustration:-Runoff-to}.}

\end{table}

\bibliographystyle{apalike}
\bibliography{Report_QE}

\end{document}